\documentclass[article,pdftex,10pt,a4paper,twocolumn]{article}
\usepackage[authoryear]{natbib}
\usepackage{graphics}

\usepackage{url}
\usepackage{epstopdf}
\usepackage{graphicx}

\RequirePackage{color}

\begin{document}

% \title{Large-scale parity violation of galaxy spin patterns}
% \title{Cosmological-scale parity violation of galaxy spin patterns}
\title{Large-scale patterns of galaxy spin rotation show cosmological-scale parity violation and multipoles}

\author{Lior Shamir \\ \small Kansas State University, Manhattan, KS 66502 \\ \small lshamir@mtu.edu}

% \author{Lior Shamir\altaffilmark{1}}
% \affil{Kansas State University \\ Manhattan, KS 66506, USA}

\maketitle

\begin{abstract}
% Spiral galaxies are unique astronomical objects in the sense that their visual appearance depends on the perspective of the observer. Since the spin patterns of spiral galaxies (clockwise or counterclockwise) are expected to be randomly distributed, in a sufficiently large universe no difference between clockwise and counterclockwise galaxies is expected.
% However, accumulating evidence start to suggest links between galaxy spin pattern at the cosmological scale. 
The distribution of spin direction of $\sim6.4\cdot10^4$ spiral galaxies with spectra was examined. The analysis shows a statistically significant cosmological-scale asymmetry between galaxies with opposite spin direction. The data also reveals that the asymmetry changes with the direction of observation, and with the redshift. The redshift dependence shows that the distribution of the spin direction of galaxies becomes more homogeneous as the redshift gets higher. The data also show photometric differences between galaxies with opposite spin patterns. When normalizing the data by the redshift, the photometric asymmetry is eliminated. However, when normalizing the data by the magnitude, statistically significant differences in the redshift remain. These evidence suggest a violation of the cosmological isotropy and homogeneity assumptions. Fitting the distribution of the galaxy spin directions to a quadrupole alignment provides fitness with statistical significance $>5\sigma$, which grows to $>8\sigma$ when just galaxies with $z>0.15$ are used. The data analysis process is fully automatic, and it is based on deterministic and symmetric algorithms with defined rules. It does not involve neither manual analysis of the data that can lead to human perceptual bias, nor machine learning that can capture human biases or other subtle differences that are difficult to identify due to the complex and non-intuitive nature of machine learning processes.
% While it is difficult to explain the observation, reasons can be related to the large-scale structure of the universe, but can also be related to the rotation of the Galaxy compared to the rotation of the observed face-on spiral galaxies. It should be noted that the dataset might be biased by the selection of spectorscopic objects.
\end{abstract}

% http://debunkingalecmacandrew.blogspot.com/2014/10/debunking-palm-and-macandrew-on-cmb_22.html

\section{Introduction}
\label{introduction}

Spiral galaxies are unique astronomical objects in the sense that their visual appearance depends on the perspective of the observer. Since the spin patterns of spiral galaxies (clockwise or counterclockwise) are expected to be randomly distributed, in a sufficiently large universe no difference between clockwise and counterclockwise galaxies is expected. However, analysis of large datasets of spiral galaxies showed photometric differences between spiral galaxies with clockwise spin patterns and spiral galaxies with counterclockwise spin patterns \citep{shamir2013color,shamir2016asymmetry,shamir2017colour,shamir2017photometric,shamir2017large}. Early attempts to identify differences between clockwise and counterclockwise galaxies did not identify statistically significant asymmetry \citep{iye1991catalog,land2008galaxy}. However, these experiments were based on much smaller datasets of just a few thousand galaxies \citep{iye1991catalog}, or on heavily biased manual classification performed by untrained volunteers \citep{land2008galaxy}. Experiments using manually annotated galaxies \citep{longo2011detection} and automatically annotated galaxies \citep{shamir2012handedness} showed statistically significant differences between the number of clockwise and counterclockwise spiral galaxies.

First evidence of photometric differences between clockwise and counterclockwise galaxies were observed using SDSS galaxies \citep{shamir2013color}. Machine learning algorithms showed accuracy much higher than mere chance in identifying the spin pattern of the galaxy by its photometric variables, showing a very strong statistically significant link between the photometry of the galaxy and its spin pattern \citep{shamir2016asymmetry}. That was shown with manually and automatically classified datasets of galaxies \citep{shamir2016asymmetry}.

Experiments with a much larger dataset of $\sim$162K automatically annotated SDSS galaxies \citep{kuminski2016computer} showed very strong statistically significant photometric differences between clockwise and counterclockwise galaxies. The experiments showed color \citep{shamir2017colour} and magnitude \citep{shamir2017photometric} differences that change with the direction of observation, and have a cosine dependence with the RA \citep{shamir2017colour,shamir2017photometric}. With SDSS data, maximum asymmetry was observed in the RA range of $(120^o,210^o)$.

Analysis of SDSS and PanSTARRS galaxies showed that data collected by both telescopes show the same asymmetry, as well as the same asymmetry pattern \citep{shamir2017large}. In both telescopes, the asymmetry changed with the direction of observation. A third dataset of $\sim$40K manually classified SDSS galaxies also showed the same pattern, with strong statistical significance of the asymmetry \citep{shamir2017large}. The observation that the asymmetry was identified by two different telescopes shows that no error in the photometric pipeline of a specific sky survey is the cause of the asymmetry. A software error is also unlikely, as such error would have exhibited itself in the form of consistent bias throughout the entire sky, while the actual observation showed very strong link between the magnitude (and direction) of the asymmetry and the direction of observation. Additionally, data that were annotated manually showed results that were in excellent agreement with the automatically annotated data \citep{shamir2017large}. In any case, it is difficult to think of a software or hardware error that would exhibit itself in the form of asymmetry between galaxies with different spin patterns. % A smaller set of $\sim$5K galaxies from COSMOS showed photometric asymmetry that agrees with the asymmetry of SDSS and PanSTARRS at around the center of the COSMOS field \citep{shamir2019photometric}.

Smaller-scale observations showed that neighboring galaxies tend to have orthogonal spin patterns, which can be explained by gravitational interactions leading to faster merging of systems of galaxies with the same direction of rotation \citep{sofue1992spins,puerari1997relative}. Other observations provide evidence of strong correlation between neighboring galaxies and their spin patterns, even in cases of galaxies that are too far from each other to have gravitational interaction \citep{lee2019mysterious}. These observations show certain evidence of patterns at the cosmological scale reflected by the distribution of galaxy spin directions \citep{lee2019mysterious}. 

Here I analyze $\sim6.4\cdot10^4$ galaxies with spectra to identify changes in the population of clockwise and counterclockwise galaxies based on their redshift. The analysis is based on SDSS DR14 galaxies with spectra.

\section{Data}
\label{data}

The initial dataset is galaxies with spectra from the SDSS DR14. All objects with class ``GALAXY'' in the SpecObjAll table were selected in the initial query, providing a set of 2,644,145 objects with spectra identified as galaxies by SDSS pipeline. Since many of the objects are too small to allow any morphological analysis, a subset of these galaxies was selected such that the Petrosian radius computed on the g band was greater than 5.5''. That selection reduced the number of galaxies with spectra to a subset of 589,049 galaxies. The distribution of the galaxies in different RA and redshift ranges is specified in Table~\ref{distribution_table}.

\begin{table}
{
%\footnotesize
\scriptsize
\begin{tabular}{|l|c|c|c|c|}
\hline
z &  0$^o$-120$^o$    &  120$^o$-240$^o$  & 240$^o$-360$^o$ & Total \\      
\hline
0-0.05   &    9,655   &   64,092  &  10,862     &    84,609    \\
0.05-0.1 &   14,746  &  104,339   & 21,098   &   140,183    \\
0.1-0.15 &    12,142  &  67,712  &  13,797   &    93,651    \\   
0.15-0.2  &  7,757    & 33,957  &  8,360     &     50,074    \\
$>$0.2  &  47,456    & 134,706    & 36,473    &   218,635    \\
\hline
Total      &  91,965 &  406,185  & 90,899   & 589,049  \\
\hline        
\end{tabular}
\caption{Distribution of the galaxies in the dataset by redshift and RA ranges.}
\label{distribution_table}
}
\end{table}

The galaxy images were fetched from SDSS Skyserver using the ``cutout'' web service, and the output images were 120$\times$120 color JPEG images. To make sure that the entire galaxy fits inside the image, if more than 25\% of the pixels on the edge of the image were bright pixels (with grayscale value greater than 125), the image was downscaled and downloaded again until the number of bright pixels was less than 25\% of the total number of pixels on the edge. The initial scale of the image was 0.1 arcseconds per pixel, and it was reduced by 0.01 arcseconds per pixel in each iteration, until the galaxy fits in the frame \citep{kuminski2016computer}.

\begin{figure*}
\includegraphics[scale=1.0]{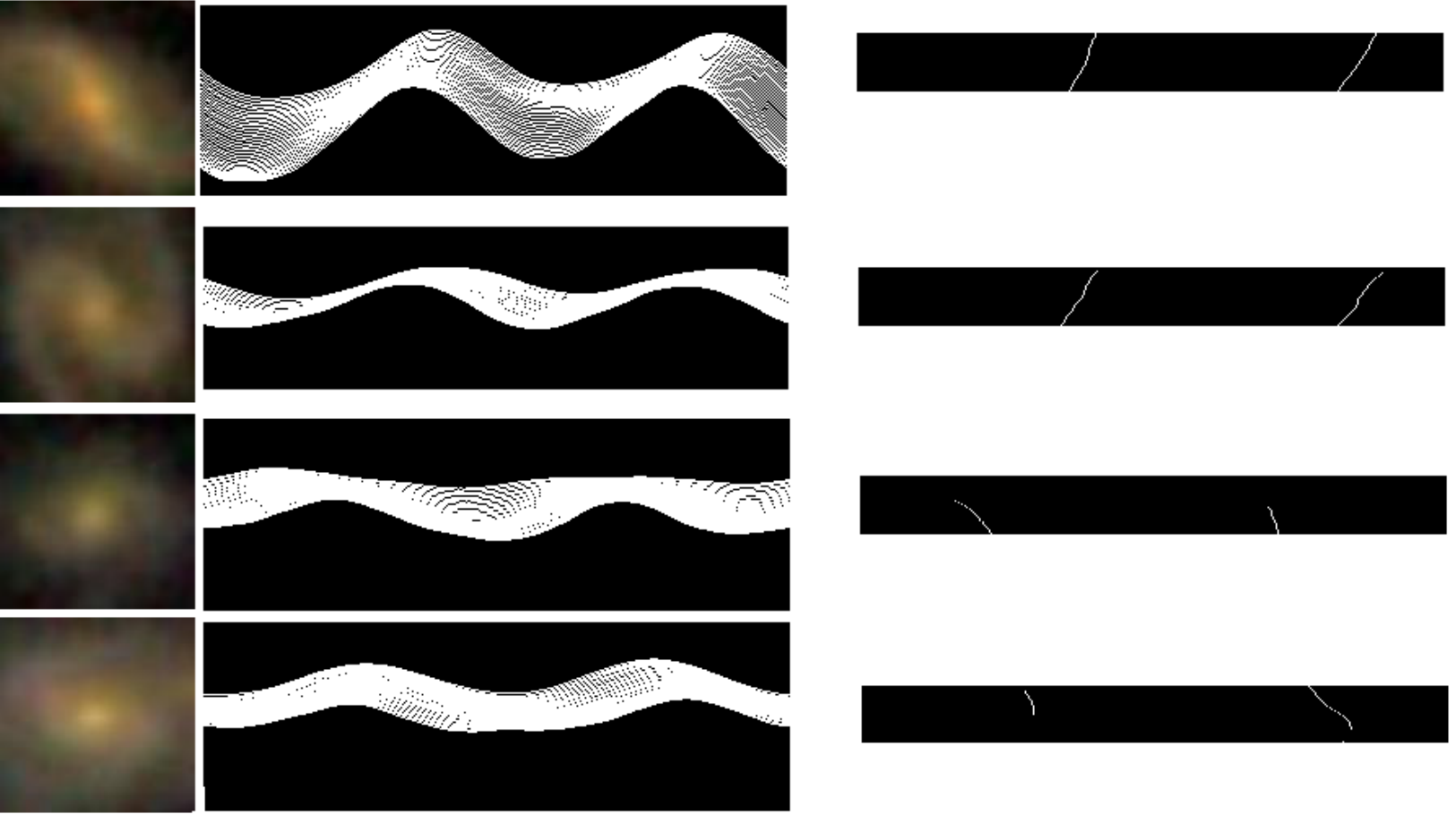}
\caption{Original images (left), their corresponding radial intensity plots (middle), and the peaks detected in the radial intensity plots (right). The sign of the lines of the peaks determines the curve of the arms of the galaxy, and consequently its direction of rotation}
\label{redial_intensity_plots}
\end{figure*}

The galaxies were classified into galaxies with clockwise spin patterns and galaxies with counterclockwise spin patterns by using the Ganalyzer tool \citep{shamir2011ganalyzer,ganalyzer_ascl}, as was done in \citep{shamir2012handedness,shamir2013color,hoehn2014characteristics,dojcsak2014quantitative,shamir2016asymmetry,shamir2017colour,shamir2017photometric,shamir2017large}. Ganalyzer is a simple deterministic algorithm that is based on converting a galaxy image into its radial intensity plot \citep{shamir2011ganalyzer}, followed by peak detection in different radial distances from the center of the galaxy. The value of the pixel $(x,y)$ in the radial intensity plot is the median value of the 5$\times$5 pixels around $(O_x+\sin(\theta) \cdot r,O_y-\cos(\theta)\cdot r)$ in the original image, where $\theta$ is the polar angle and {\it r} is the radial distance. Because the arm of the galaxy is expected to be brighter than other parts of the galaxy at the same distance from the galaxy center, the groups of peaks in the radial intensity plot are expected to be associated with the arms of the galaxy. Figure~\ref{redial_intensity_plots} shows examples of original galaxy images, their corresponding radial intensity plots, and the peaks detected in the radial intensity plots.

The sign of the linear regression of the peaks determines the curve of the arms, and consequently the rotation direction of the galaxy. Obviously, not all galaxies are spiral, and not all spiral galaxies have identifiable spin patterns. To have good identification of the direction of the peaks, at least 30 peaks need to be detected in the radial intensity plot of the galaxy, otherwise the galaxy is classified as unidentifiable and rejected from the analysis. The algorithm is described in detail in \citep{shamir2011ganalyzer}, as well as in \citep{shamir2012handedness,shamir2013color,hoehn2014characteristics,dojcsak2014quantitative,shamir2016asymmetry,shamir2017colour,shamir2017photometric}.

An important advantage of Ganalyzer is that it is a model-driven algorithm that is not based on machine learning. Therefore, human bias or other learning bias such as the part of the sky from which the training samples were taken cannot affect the performance or behavior of the algorithm. Ganalyzer is a straightforward open source deterministic algorithm that is not based on complex non-intuitive rules often used by machine learning algorithms, especially with deep neural networks. Ganalyzer is therefore not subjected to any training bias, as no training is required in any stage. The rules are designed in a fully symmetric manner, so that no bias of the algorithm to a certain type of spin pattern is allowed by the code.

\begin{figure}[h]
\includegraphics[scale=0.62]{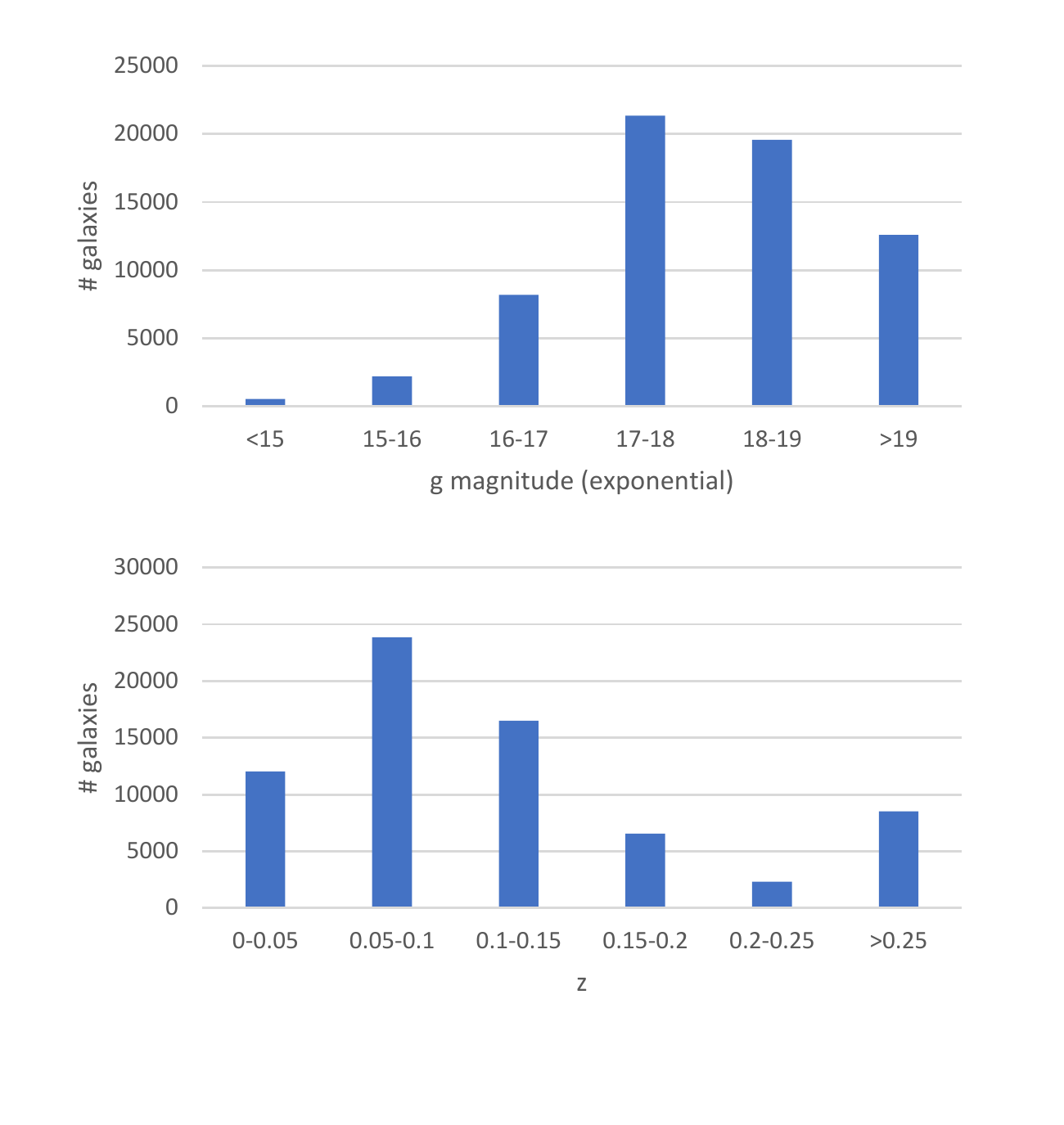}
\caption{The distribution of the exponential magnitude (g) and redshift of the galaxies used in the dataset.}
\label{distribution}
\end{figure}

The process resulted in a dataset of 32,055 galaxies with clockwise spin patterns, and 32,501 galaxies with counterclockwise spin patterns. Visual inspection of 100 randomly selected galaxies showed no cases of missclassified galaxies, indicating that the dataset is reasonably clean. SDSS can sometimes assign more than one object ID to the same galaxy. Removing these duplication provided a dataset of 63,693 galaxies, 31,666 clockwise galaxies and 32,027 are counterclockwise. The color and redshift distribution of the entire dataset population is shown in Figure~\ref{distribution}.

%  select count(*) from MyDB.DR14_ccw ,specObjAll where DR14_ccw.ID=specObjAll.bestObjID and specObjAll.z>0.2 and specObjAll.z<=0.5

\begin{figure}[h]
\includegraphics[scale=0.60]{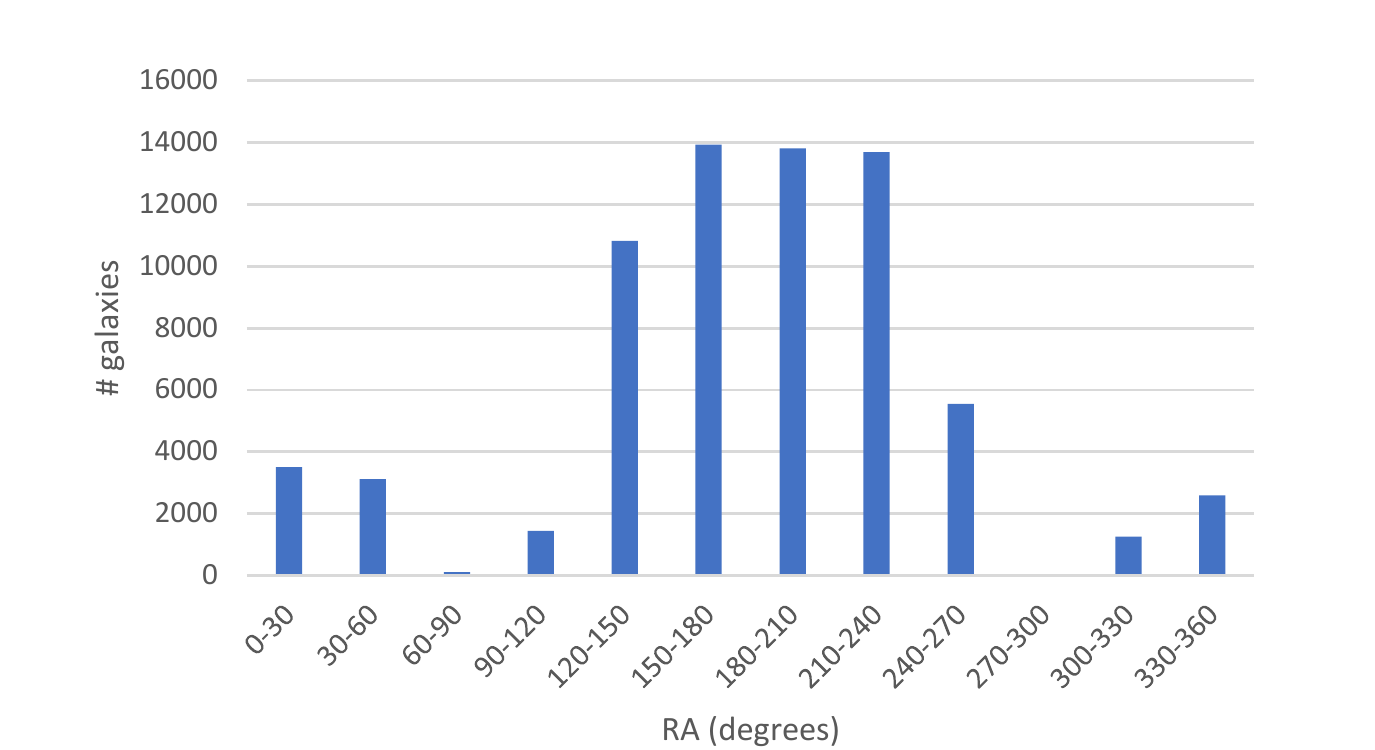}
\caption{The distribution of the galaxies in the different RA ranges.}
\label{distribution_ra}
\end{figure}

The galaxies with spectroscopy are not distributed uniformly in the sky covered by SDSS, and some parts of the sky contain more spectroscopic objects than others. Figure~\ref{distribution_ra} shows the number of galaxies in each RA range. As the graph shows, the number of galaxies with spectra is particularly low in the RA ranges of $(60^o,90^o)$ and $(270^o,300^o)$.  

\section{Results}
\label{results}

Following the observations described in \citep{shamir2017large}, the galaxies were divided into four groups based on their right ascension. The division of the sky into regions is based on the observation that the asymmetry between clockwise and counterclockwise galaxies depends on the direction of observation \citep{shamir2017colour,shamir2017photometric,shamir2017large}, and the asymmetry is strongest in the RA range of (120$^o$,210$^o$). Mirroring the galaxy images and repeating the same analysis leads to the exact opposite results, which is expected since Ganalyzer is a symmetric and deterministic algorithm, and therefore the annotation of a certain galaxy image is the opposite of the annotation of its mirrored image. Table~\ref{directions} shows the number of clockwise and counterclockwise galaxies in each direction of observation.

% select count(*) from MyDB.DR14_ccw ,specObjAll where DR14_ccw_rad5_5.ID=specObjAll.bestObjID and specObjAll.z>0.05 and specObjAll.z<=0.1 and specObjAll.ra>120 and specObjAll.ra<210

% select count(ID) from MyDB.DR14_cw ,PhotoObjAll where DR14_cw.ID=PhotoObjAll.objID and (ra>300 or ra<30)

\begin{table}
{
%\footnotesize
\scriptsize
\begin{tabular}{|l|c|c|c|c|c|}
\hline
RA &    cw    &  ccw  & ${cw}\over{cw+ccw}$ & P value & q-value \\      
\hline
%120$^o$-210$^o$     & 18110 & 18636 & 0.493    & 0.0029  & 0.0186  \\
120$^o$-210$^o$     & 17876 & 18391 & 0.493    & 0.0034  & 0.0136  \\

%$ > 300^o < 30^o$   & 3099 & 3258    & 0.487    & 0.022    & 0.089 \\
$>300^o<30^o$   & 3064 & 3219    & 0.487    & 0.026    & 0.104 \\

%30$^o$-120$^o$      & 1694 & 1625     & 0.51      &	0.11  & 0.45 \\
30$^o$-120$^o$      & 1665 & 1560     & 0.516      &	0.033  & 0.132 \\

%210$^o$-300$^o$    & 9152 & 8982     &  0.504   & 0.1        & 0.41  \\
210$^o$-300$^o$    & 9061 & 8857     &  0.506   & 0.065        & 0.258  \\

\hline
\end{tabular}
\caption{The distribution of the clockwise and counterclockwise galaxies in different RA ranges. The P value of the binomial distribution shows the probability that the asymmetry between the number of clockwise and counterclockwise galaxies occurs by chance. The q-value is the Bonferroni-corrected P value.}
\label{directions}
}
\end{table}

The table shows statistically significant difference between the number of clockwise galaxies and the number of counterclockwise galaxies in the RA range of $(120^o,210^o)$, the same RA range that showed photometric asymmetry between clockwise and counterclockwise galaxies \citep{shamir2017large}. The statistical power of the parity violation in that RA range remains significant even after applying the Bonferonni correction. The same RA range in the opposite hemisphere ($>300^o , <30^o$) also shows higher population of counterclockwise galaxies, but when applying the Bonferonni correction the statistical significance drops below a discovery level. It should be noted that the RA range ($ > 300^o < 30^o$) has a much lower number of galaxies, which can have a major impact on the lower statistical significance. The other two RA ranges show higher numbers of clockwise galaxies, but the statistical significance of the difference is lower.     % The strongest asymmetry observed in the RA range $(120^o,210^o)$ agrees with the strongest photometric asymmetry between clockwise and counterclockwise galaxies, which also peaked at that RA range \citep{shamir2017colour,shamir2017photometric,shamir2017large}.

Table~\ref{directions} also shows agreement in the direction of asymmetry between parts of the sky in opposite hemispheres. The direction of asymmetry in the RA range (120$^o$,210$^o$) agrees with the direction of asymmetry in ($>300^o<30^o$), as in both parts of the sky the number of counterclockwise galaxies is greater than the number of clockwise galaxies. In the RA ranges (30$^o$,120$^o$) and (210$^o$,300$^o$) there is an excessive number of clockwise galaxies. That provide certain evidence to the existence of a quadrupole. Assuming a quadrupole, the RA ranges (120$^o$,210$^o$) and ($>300^o<30^o$) have a combined total of 42,550 galaxies, such that 21,610 of them have counterclockwise spin. Assuming random spin patterns, the probability of such distribution by chance is $(P<0.0006)$. In the RA ranges (120$^o$,210$^o$) and ($>300^o<30^o$) the total number of galaxies is 21,143, and 10,726 have clockwise spin patterns. The probability for such distribution by mere chance is $(P<0.017)$.

Table~\ref{dis_120_210} shows the distribution of clockwise and counterclockwise galaxies in the RA range of $(120^o,210^o)$ and different redshift ranges. The table shows the number of clockwise and counterclockwise galaxies in each redshift range. The P value shows the probability to have such distribution by chance, as determined by the accumulative binomial distribution assuming that the probability of a galaxy to have a clockwise or counterclockwise spin is exactly 0.5. As the table shows, the asymmetry between the number of clockwise and counterclockwise galaxies increases with the redshift. While the asymmetry between clockwise and counterclockwise galaxies is insignificant in redshift range of 0-0.05, the asymmetry grows consistently with the redshift range. That cannot be explained by changes in the performance of the galaxy classification algorithm, as lower performance of the algorithm should shift the difference closer to random distribution, which is 0.5, and in any case should have also affected the other RA ranges.

\begin{table}
{
%\footnotesize
\scriptsize
\begin{tabular}{|l|c|c|c|c|}
\hline
z &    cw    &  ccw  & ${cw}\over{cw+ccw}$ & P value \\      
\hline
%0-0.05     &	3426 &	 3384 & 	0.5003	& 0.698  \\
0-0.05     &	3216 &	 3180 & 	0.5003	& 0.698  \\

%0.05-0.1 &	6631	& 6661 &	0.498 &	0.4 \\
0.05-0.1 &	6240	& 6270 &	0.498 &	0.4 \\

%0.1-0.15 &	4503 &	4558 & 	0.496 &	0.285  \\
0.1-0.15 &	4236 &	4273 & 	0.496 &	0.285  \\

%0.15-0.2 &	1682 &	1825 & 	0.479 &	0.008  \\
0.15-0.2 &	1586 &	1716 & 	0.479 &	0.008  \\

%0.2 - &	2500 &	 2822 &	 0.469 &	$5.37\cdot10^{-6}$  \\
0.2 - 0.5 &	2598 &	 2952 &	 0.469 &	$1.07\cdot10^{-6}$  \\

\hline
Total & 17,876 & 18,391 & 0.493 & 0.0034 \\
\hline
\end{tabular}
\caption{The distribution of the clockwise and counterclockwise galaxies in redshift ranges in the RA range of $(120^o,210^o)$. The P value of the accumulative binomial distribution shows the probability that the asymmetry between the number of clockwise and counterclockwise galaxies occurs by chance.}
\label{dis_120_210}
}
\end{table}

The Pearson correlation between each of the five redshift ranges and the asymmetry in each redshift range is $\sim$-0.951 (P$<$0.013). The Pearson correlation can also be computed without separating the dataset into redshift ranges. Assigning all clockwise galaxies to 1 and all counterclockwise galaxies to -1 provides a sequence of 63,693 pairs of values such that the redhisft of each galaxy is paired with 1 or -1, based on the spin direction of the galaxy. The Pearson correlation between these pairs of values is -0.01546. The correlation coefficient is low, as expected, but when the size of the dataset is 63,693 pairs the probability of having such correlation by chance is (P$\simeq$0.00009). The correlation between the rotation direction (1 or -1) and the redshift in the RA range $(120^o,210^o)$ is -0.02329. With the 36,267 galaxies in that RA range the probability to get such correlation by chance is (P$<$0.0001). That shows that in the RA range $(120^o,210^o)$ the population of counterclockwise galaxies observed by SDSS increases with the redshift.

% The Pearson correlation between the rotation direction (1 or -1) and redshift in all four RA ranges is shown in Table~\ref{pearson_all}.

% \begin{table}
% {
% %\footnotesize
% \scriptsize
% \begin{tabular}{|l|c|c|}
% \hline
% RA &    r    &  P  &  \\      
% \hline
% 120$^o$-210$^o$   &  -0.02329 & $<$0.0001   \\
% $ > 300^o < 30^o$ & -0.0178 \\
% 30$^o$-120$^o$ & 0.0233 & 0.18 \\
% 210$^o$-300$^o$ & -0.007 &   0.34 \\
% All & -0.01546 & $<$0.0001 \\
% \hline
% \end{tabular}
% \caption{Pearson correlation (r) and P value of the correlation in the different RA ranges}
% }
% \end{table}

The number of clockwise and counterclockwise galaxies in each redshift range in the three other RA ranges are specified in Tables~\ref{dis_30_120} through~\ref{dis_30_300}. 

\begin{table}
{
%\footnotesize
\scriptsize
\begin{tabular}{|l|c|c|c|c|}
\hline
z &    cw    &  ccw  & ${cw}\over{cw+ccw}$ & P value \\      
\hline
% 0-0.05     &	380 &	 457 & 0.454	& 0.004  \\
0-0.05     &	245 &	 260 & 0.485	& 0.266  \\
%0.05-0.1 &	569	& 580 & 0.495 &	0.384 \\
0.05-0.1 &	442	& 429 & 0.507 &	0.34 \\
% 0.1-0.15 &	575 &	578 & 	0.498 &	0.476  \\
0.1-0.15 &	370 &	384 & 	0.491 &	0.32  \\
% 0.15-0.2 &	299 &	241 & 	0.553 &	0.007  \\
0.15-0.2 &	 229 &	185 & 	0.553 &	0.017  \\
0.2 - 0.5 &	 379 &	 302 &	 0.556 &	0.0018  \\
\hline
Total  & 1,665 & 1,560 & 0.516 & 0.034 \\
\hline
\end{tabular}
\caption{The distribution of the clockwise and counterclockwise galaxies in different redshift ranges in the RA range of $(30^o,120^o)$.}
\label{dis_30_120}
}
\end{table}

\begin{table}
{
%\footnotesize
\scriptsize
\begin{tabular}{|l|c|c|c|c|}
\hline
z &    cw    &  ccw  & ${cw}\over{cw+ccw}$ & P value \\      
\hline
%0-0.05     &	1749 & 1743 & 0.5	& 0.466  \\
0-0.05     &	1643 & 1635 & 0.501	& 0.451  \\

%0.05-0.1 &	3575	& 3410 & 0.511 &	0.025 \\
0.05-0.1 &	3351	& 3162 & 0.514 &	0.01 \\

%0.1-0.15 &	2274 & 2300 & 0.497 &	0.355  \\
0.1-0.15 &	2105 & 2115 & 0.499 &	0.445  \\

%0.15-0.2 &	855 &	851 & 	0.501 &	0.471  \\
0.15-0.2 &	797 &	782 & 	0.500 &	0.362  \\

%0.2 - &	1251 & 1243 &  0.501 &	0.444  \\
0.2 - 0.5 &	1165 & 1163 &  0.500 &	0.492  \\
\hline
Total  & 9,061 &  8,857 & 0.505 & 0.07 \\
\hline
\end{tabular}
\caption{The distribution of the clockwise and counterclockwise galaxies in different redshift ranges in the RA range of $(210^o,300^o)$.}
\label{dis_210_300}
}
\end{table}

\begin{table}
{
%\footnotesize
\scriptsize
\begin{tabular}{|l|c|c|c|c|}
\hline
z &    cw    &  ccw  & ${cw}\over{cw+ccw}$ & P value \\      
\hline
% 0-0.05     &	421 &	 477 & 0.469	& 0.033  \\
0-0.05     &	360 &	 402 & 0.472	& 0.068  \\

%0.05-0.1 &	1248	& 1205 & 0.508 &	0.198 \\
0.05-0.1 &	1040	& 1023 & 0.504 &	0.362 \\

%0.1-0.15 &	835 &	874 & 	0.488 &	0.178  \\
0.1-0.15 &	714 &	735 & 	0.492 &	0.299  \\

%0.15-0.2 &	364 &	432 & 	0.457 &	0.008  \\
0.15-0.2 &	322 &	379 & 	0.459 &	0.017  \\

%0.2 - &	709 &	 793 &	 0.472 &	0.016  \\
0.2 -0.5 &	628 &	 680 &	 0.48 &	0.079  \\

\hline
Total  & 3,064 &  3,219 & 0.487 & 0.026 \\
\hline
\end{tabular}
\caption{The distribution of the clockwise and counterclockwise galaxies in different redshift ranges in the RA range of $(<30^o, >300^o)$.}
\label{dis_30_300}
}
\end{table}

The difference between the population of clockwise and counterclockwise galaxies in the different RA and redshift ranges is shown in Table~\ref{asymmetry_all}. The table shows the RA ranges in thinner slices of 30$^o$, and the redshift range slice is 0.15. The RA ranges of (60$^o$,90$^o$) and (270$^o$,300$^o$) are ignored due to the very low number of galaxies in them of 83 and 0, respectively. The error is determined by $\frac{1}{\sqrt{n}}$, such that n is the number of galaxies. Naturally, due to the separation of the data into many smaller sections, in most sections the number of galaxies does not allow statistical significance. However, the RA range of (120$^o$,180$^o$) shows statistically significant asymmetry between the population of clockwise and counterclockwise galaxies.

\begin{table*}
{
% \footnotesize
% \scriptsize
\tiny
\begin{tabular}{|l|c|c|c|c|c|c|c|c|c|c|}
\hline
z &    0-30    &  30-60  & 90-120 & 120-150 & 150-180 & 180-210 & 210-240 & 240-270 & 300-330 & 330-360  \\      
\hline
0-0.15     &   -0.046$\pm$0.02	& -0.035$\pm$0.03   & 0.027$\pm$0.03  & 0.014$\pm$0.01  & -0.017$\pm$0.01  & 0.004$\pm$0.01  & 0.01$\pm$0.01  & 0.021$\pm$0.02 & 0.03$\pm$0.03 & 0.014$\pm$0.03    \\
0.15-0.3  &  -0.114$\pm$0.04 &	0.072$\pm$0.04 &	-0.031$\pm$0.07 &	-0.073$\pm$0.02 &	-0.053$\pm$0.02 &	0.03$\pm$0.02 &  0.035$\pm$0.02	& -0.045$\pm$0.04 & -0.086$\pm$0.07 & 0.007$\pm$0.04    \\
% $>$0.3   &	  0.031$\pm$0.05   &	0.11$\pm$0.06   &		0.002$\pm$0.09   &	-0.131$\pm$0.03   &	-0.156$\pm$0.03   & 	-0.012$\pm$0.03   &	0.002$\pm$0.03   &	0.007$\pm$0.05   &	-0.052$\pm$0.13   &	-0.041$\pm$0.06       \\
Total     &	 -0.051$\pm$0.02	& 0.016$\pm$0.02	&	0.017$\pm$0.03 &	-0.015$\pm$0.01 &	-0.034$\pm$0.01 &	0.006$\pm$0.01 &	0.013$\pm$0.01 &	0.007$\pm$0.01	&	0.002$\pm$0.03 &	-0.003$\pm$0.02      \\
\hline
\end{tabular}
\caption{Asymmetry (cw-ccw)/(cw+ccw) in different RA and redshift ranges.}
\label{asymmetry_all}
}
\end{table*}

% 0-0.15     &   -0.046$\pm$0.023	& -0.035$\pm$0.030   & 0.027$\pm$0.032  & 0.014$\pm$0.012  & -0.017$\pm$0.01  & 0.004$\pm$0.01  & 0.01$\pm$0.01  & 0.021$\pm$0.016 & 0.03$\pm$0.035 & 0.014$\pm$0.026    \\
% 0.15-0.3  &  -0.114$\pm$0.040 &	0.072$\pm$0.042 &	-0.031$\pm$0.072 &	-0.073$\pm$0.025 &	-0.053$\pm$0.023 &	0.03$\pm$0.023 &  0.035$\pm$0.024	& -0.045$\pm$0.037 & -0.086$\pm$0.067 & 0.007$\pm$0.047    \\
% $>$0.3   &	  0.031$\pm$0.049   &	0.11$\pm$0.065   &		0.002$\pm$0.089   &	-0.131$\pm$0.031   &	-0.156$\pm$0.028   & 	-0.012$\pm$0.028   &	0.002$\pm$0.031   &	0.007$\pm$0.047   &	-0.052$\pm$0.126   &	-0.041$\pm$0.063       \\
% Total     &	 -0.051$\pm$0.018	& 0.016$\pm$0.023	&	0.017$\pm$0.027 &	-0.015$\pm$0.01 &	-0.034$\pm$0.009 &	0.006$\pm$0.009 &	0.013$\pm$0.009 &	0.007$\pm$0.014	&	0.002$\pm$0.03 &	-0.003$\pm$0.021      \\

\begin{figure}[h]
\includegraphics[scale=0.5]{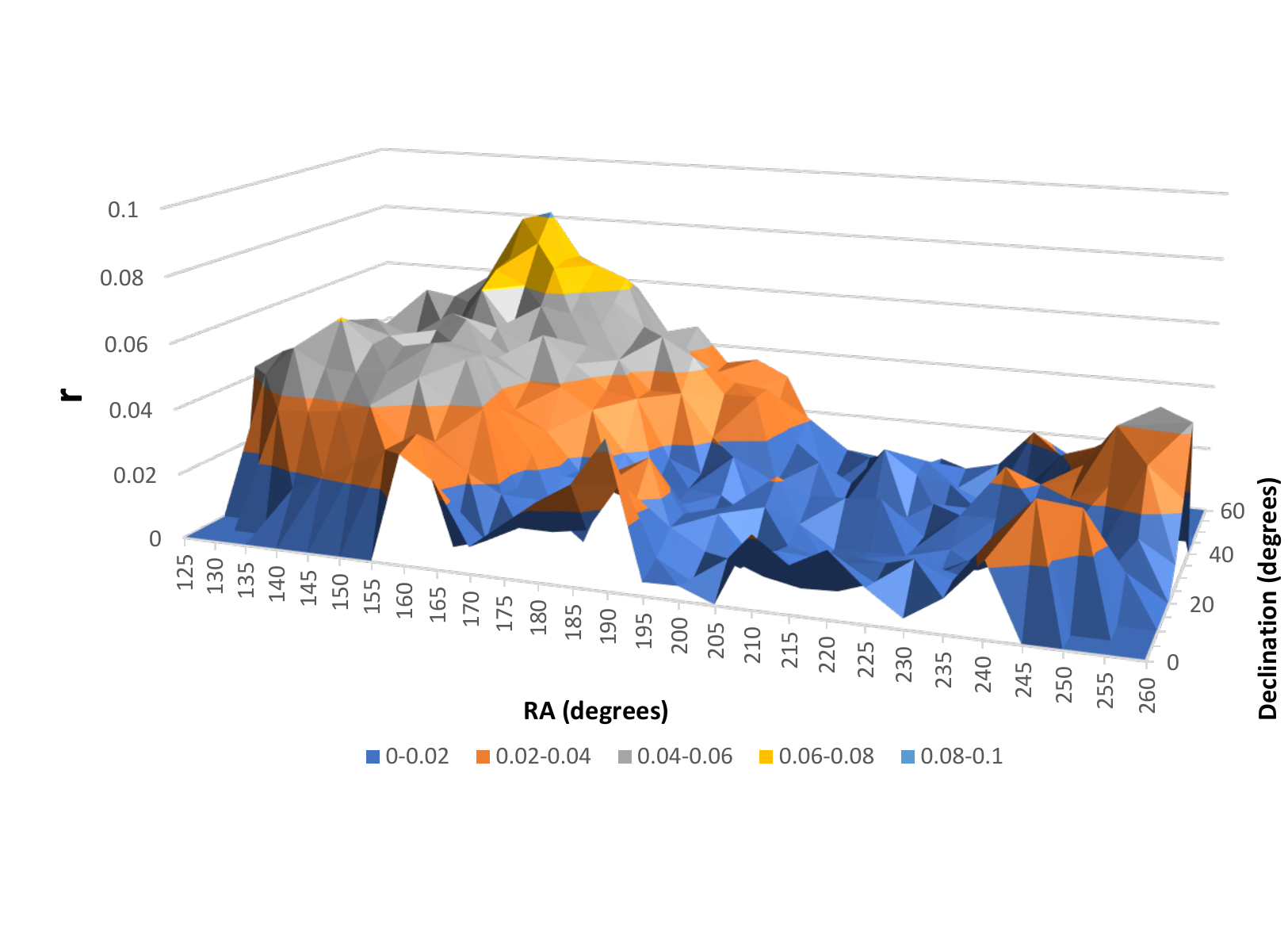}
\caption{The absolute value of the correlation between the redshift and the galaxy spin patterns in the 15$^o$ around different $(\alpha,\delta)$ combinations.}
\label{pearson_axis}
\end{figure}

To further analyze the correlation between the redshift and the rotation direction, counterclockwise galaxies were assigned the value 1, and clockwise galaxies were assigned the value -1. Then, for every possible $(\alpha,\delta)$ combination (in increments of 5), the Pearson correlation between the galaxy spin pattern (1 or -1) and the redshift was computed for all galaxies that are 15$^o$ or less away from the $(\alpha,\delta)$ coordinates. In case the $(\alpha,\delta)$ coordinates had less than 3000 galaxies within 15$^o$ or less, the $(\alpha,\delta)$ coordinates were excluded. Figure~\ref{pearson_axis} shows the correlations in difference $(\alpha,\delta)$. The strongest correlation of 0.0815 (P$<$0.00001) was identified in (160,$^o$,50$^o$). The 1$\sigma$ range for the RA is (130$^o$,185$^o$), and for the declination it is (15$^o$,65$^o$). That, however, is limited to the sky covered by SDSS, as it is not possible to measure the correlation between the redshift and the rotation of direction of galaxies in unpopulated or underpopulated sky regions.  

The asymmetry between the number of clockwise and counterclockwise galaxies changes with the direction of observation. A cosmological-scale dipole axis is expected to be observed in the form of cosine dependence with the direction of observation \citep{longo2011detection,shamir2012handedness}. To identify the most likely dipole axis, for each $(\alpha,\delta)$ combination, the galaxies were fitted into $\cos(\phi)$, such that $\phi$ is the angular distance between the $(\alpha,\delta)$ coordinates and the coordinates of the galaxy, as was done in \citep{shamir2012handedness}. That was done by assigning each galaxy with a random number within the set \{-1,1\}, and fitting $d\cdot\cos(\phi)$ to $\cos(\phi)$, such that $d$ is the randomly assigned spin direction (1 or -1). The $\chi^2$ was computed 2000 times for each $(\alpha,\delta)$ combination, and the mean and $\sigma$ were computed for each $(\alpha,\delta)$ combination. Then, the $\chi^2$ mean computed with the random spin directions was compared to the $\chi^2$ when $d$ was assigned to the actual spin direction of the galaxies. The difference (in terms of $\sigma$) between the $\chi^2$ of the actual spin patterns and the mean $\chi^2$ determined using randomly assigned spin patterns show the statistical likelihood of an axis at the $(\alpha,\delta)$ coordinates.

\begin{figure}[h]
\includegraphics[scale=0.60]{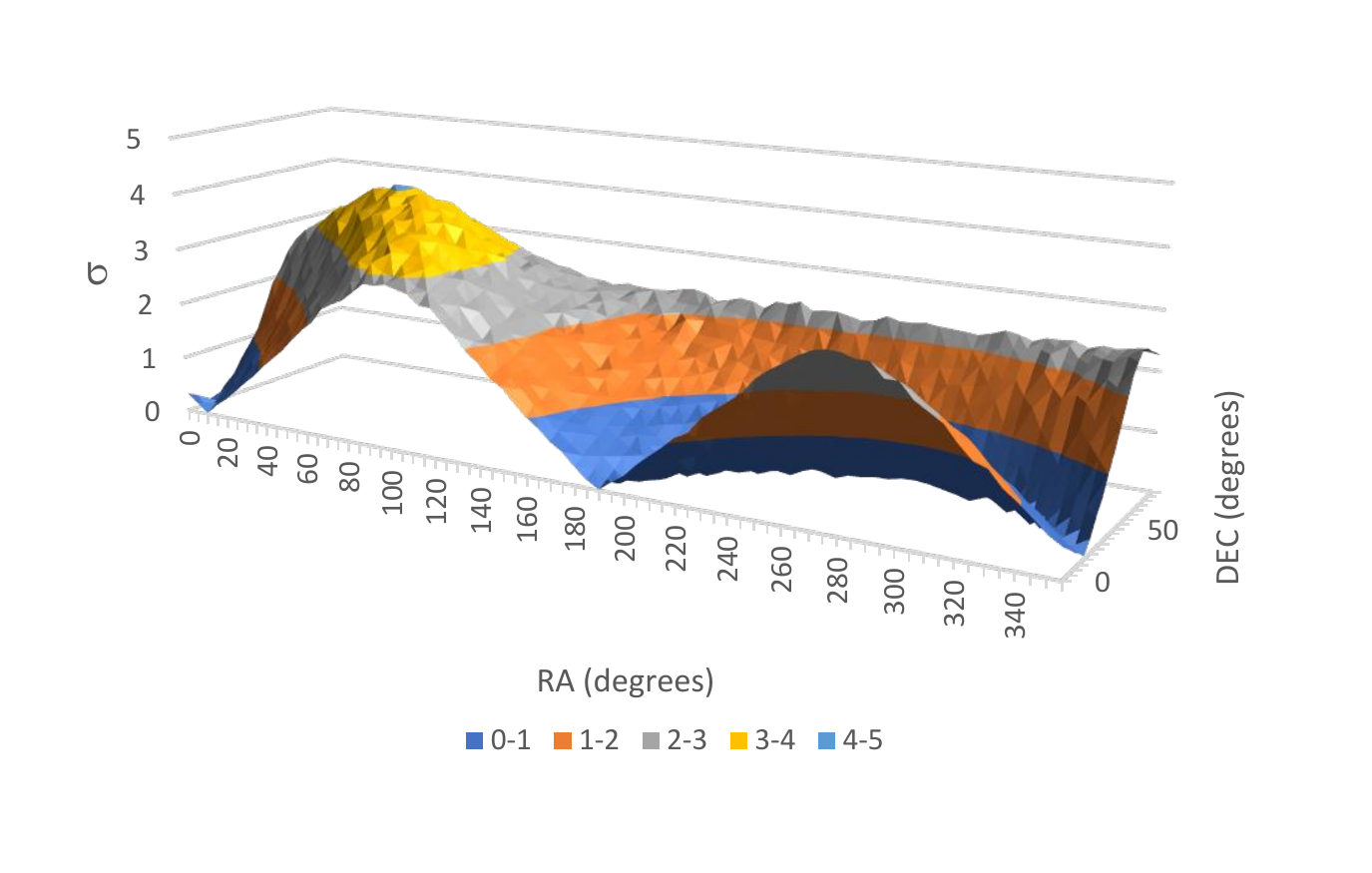}
\caption{The $\sigma$ of the likelihood of a dipole axis in different $(\alpha,\delta)$ combinations.}
\label{axis}
\end{figure}

Figure~\ref{axis} shows the $\sigma$ for the asymmetry axis of all possible integer $(\alpha,\delta)$ combinations, and for different redshift ranges. The most likely $(\alpha,\delta)$ was identified at $(\alpha=74^o,\delta=53^o)$, with $\sigma$ of $\sim$4.03, meaning that the probability of such axis to occur by chance if the rotation directions of the galaxies are random is $(P<0.00006)$. The 1$\sigma$ error is $(24^o,195^o)$ for the right ascension, and $(14^o,90^o)$ for the declination. As Table~\ref{dis_120_210} shows, the asymmetry changes across different redshift ranges. When identifying the most likely axis using just galaxies with $z>0.15$, the most likely axis is at $(\alpha=87^o,\delta=56^o)$, with a high $\sigma$ of $\sim$6.1. The 1$\sigma$ error is $(45^o,145^o)$ for the RA, and $(34^o,90^o)$ for the declination. However, when the galaxies are limited to $0<z<0.15$, the most likely axis is identified at $(57^o,32^o)$, with 1.9$\sigma$.

\begin{figure}[h]
\includegraphics[scale=0.60]{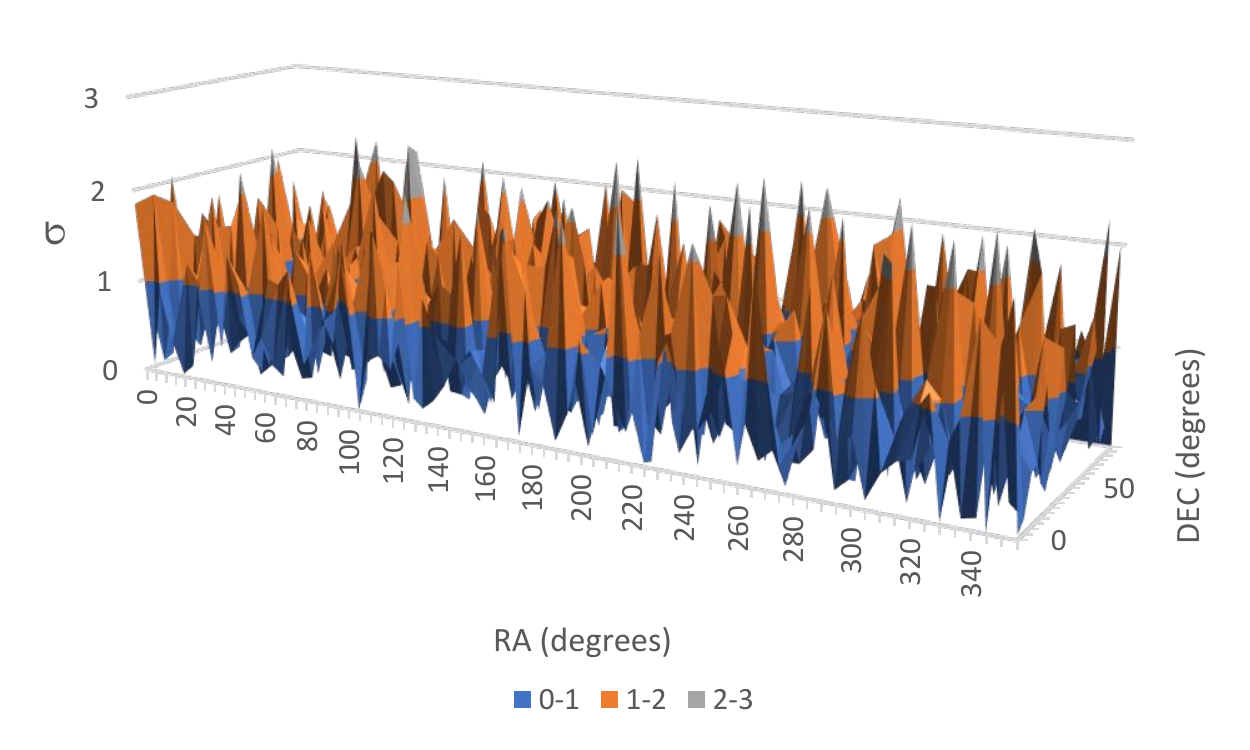}
\caption{The $\sigma$ of the axis of asymmetry in different $(\alpha,\delta)$ combinations such that galaxies were assigned with spin patterns randomly.}
\label{axis_random}
\end{figure}

Figure~\ref{axis_random} shows the result of the same experiment, but instead of using the spin patterns determined by Ganalyzer as described in Section~\ref{data}, each galaxy was assigned a random spin pattern (1 or -1). As expected, the graph does not show a specific pattern leading to an axis that can be considered the most likely axis. None of the possible $(\alpha,\delta)$ combinations showed statistical significance higher than 2.3$\sigma$.

According to Table~\ref{directions}, the asymmetry shows more counterclockwise galaxies in RA ranges $(120^o,210^o)$ and $(>300^o<30^o)$, and an excessive number of clockwise galaxies in the RA ranges of $(30^o-120^o)$ and $(210^o-300^o)$. That might be considered as certain evidence of a quadrupole alignment. Quadrupole alignment has also been observed in the context of CMB anisotropy as observed by COBE, WMAP, and Planck \citep{cline2003does,gordon2004low,zhe2015quadrupole}, suggesting the model of double inflation \citep{feng2003double,piao2004suppressing} and cosmological models that do not conform to the standard cosmology models \citep{rodrigues2008anisotropic,piao2005possible,jimenez2007cosmology}.

\begin{figure}[h]
\includegraphics[scale=0.60]{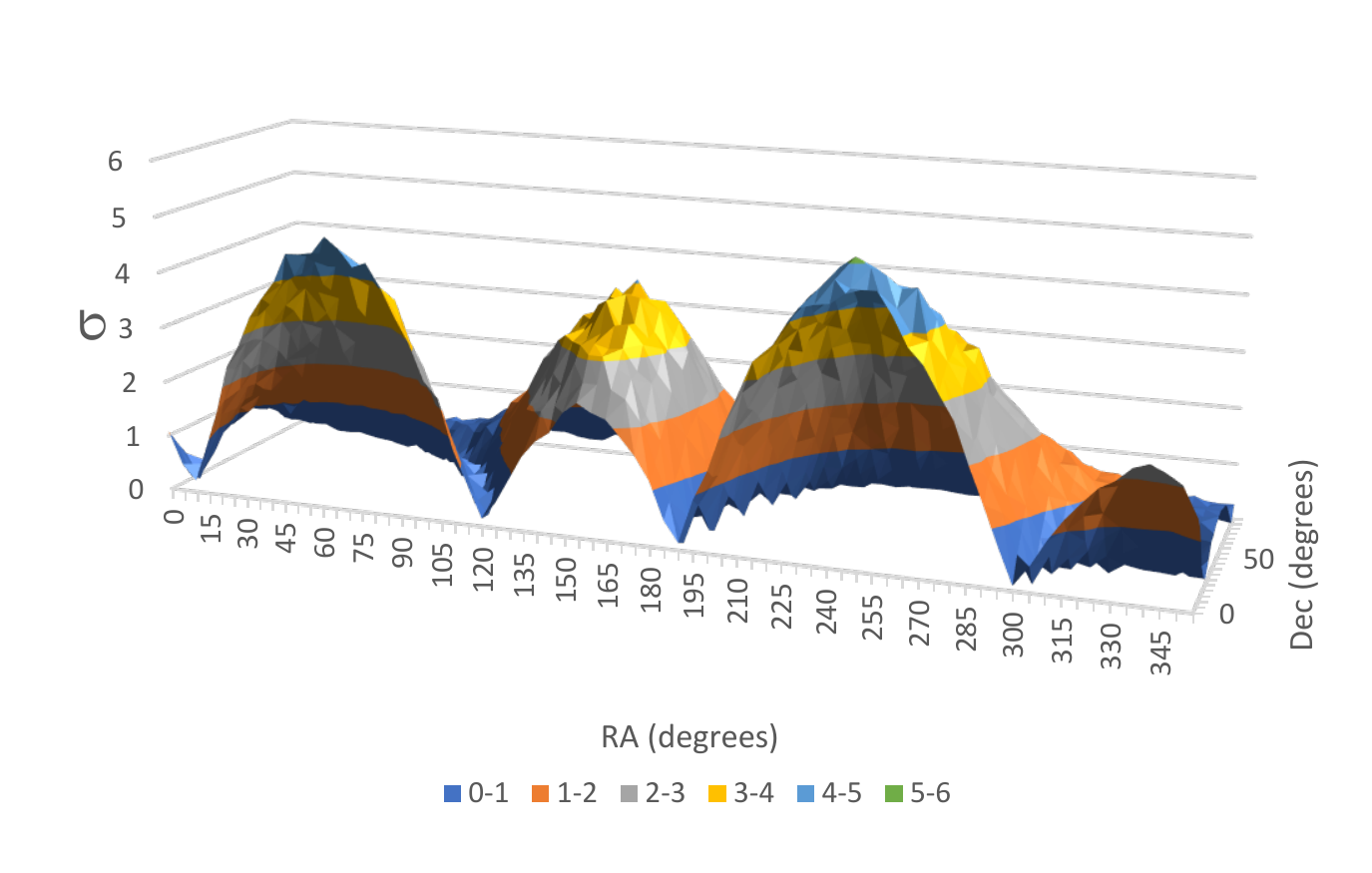}
\caption{The $\sigma$ of the likelihood of a quadrupole axis in different $(\alpha,\delta)$ combinations.}
\label{axis_quadrupole}
\end{figure}

Figure~\ref{axis_quadrupole} shows the $\chi^2$ fitting to cosine $2\phi$ dependence in each possible integer $(\alpha,\delta)$ combinations. The most likely axis is identified in $(\alpha=248^o,\delta=10^o)$, with certainty of $5.1\sigma$. The 1$\sigma$ error range of the RA is $(227^o,271^o)$, and on the declination the error range is $(0^o,33^o)$. Another axis peaks at $(\alpha=163^o,\delta=42^o)$ with 4.06$\sigma$. That axis is close to the axis of highest correlation between redshift and the asymmetry between clockwise and counterclockwise galaxies as shown by Figure~\ref{pearson_axis}. Repeating the experiment by assigning the galaxies with random spin directions provided a profile similar to the profile displayed in Figure~\ref{axis_random}.

\begin{figure}[h]
\includegraphics[scale=0.60]{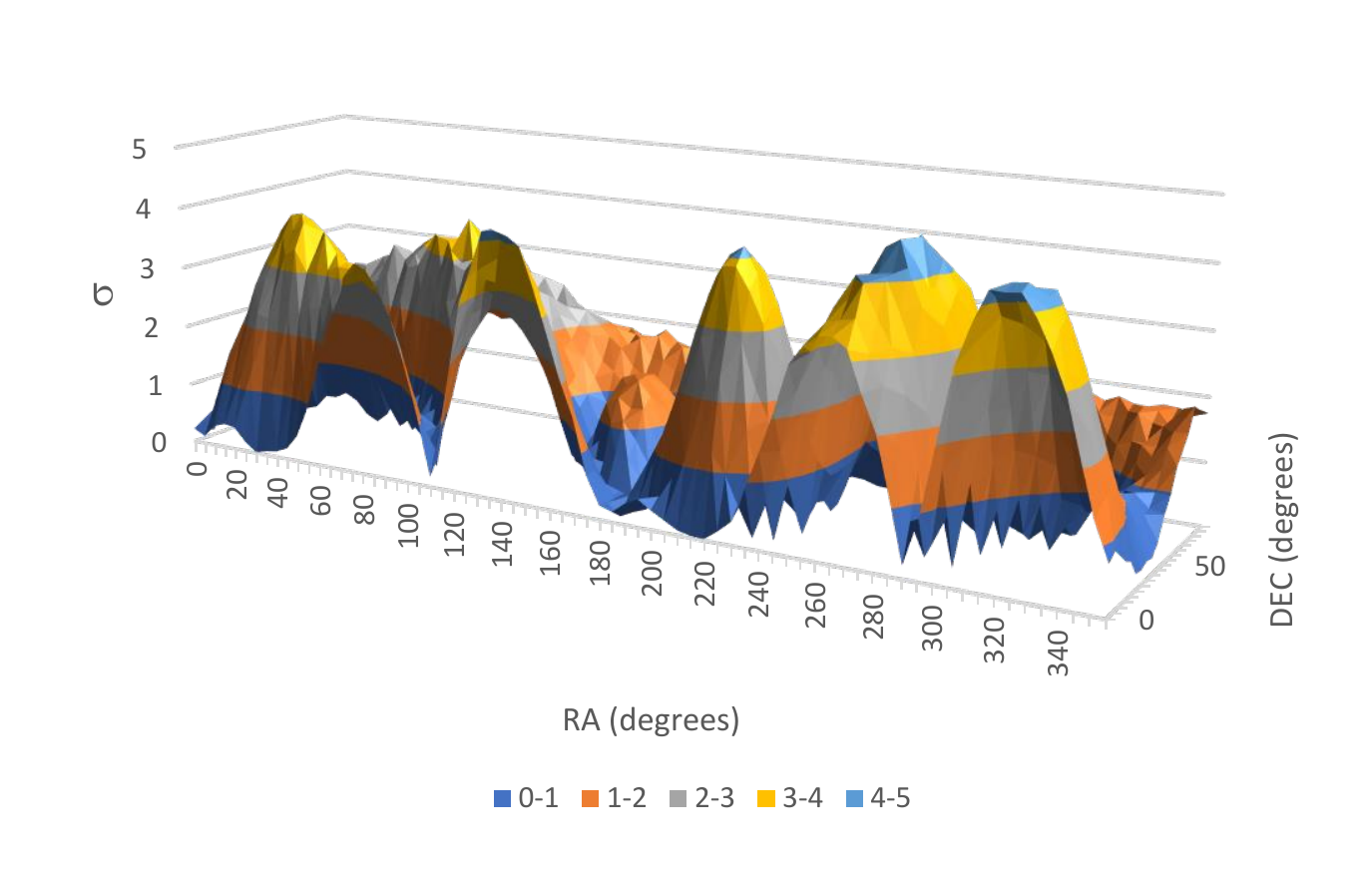}
\caption{The $\sigma$ of the likelihood of a octopole axis in different $(\alpha,\delta)$ combinations.}
\label{axis_octopole}
\end{figure}

An attempt to fit the spin directions of the galaxies to octopole alignment is displayed in Figure~\ref{axis_octopole}. The most likely axis is identified at $(\alpha=219^o,\delta=30^o)$ with likelihood of $\sigma\simeq4.16$. That likelihood is lower than the likelihood of the most likely axis when fitting to a quadrupole alignment.

\begin{figure}[h]
\includegraphics[scale=0.60]{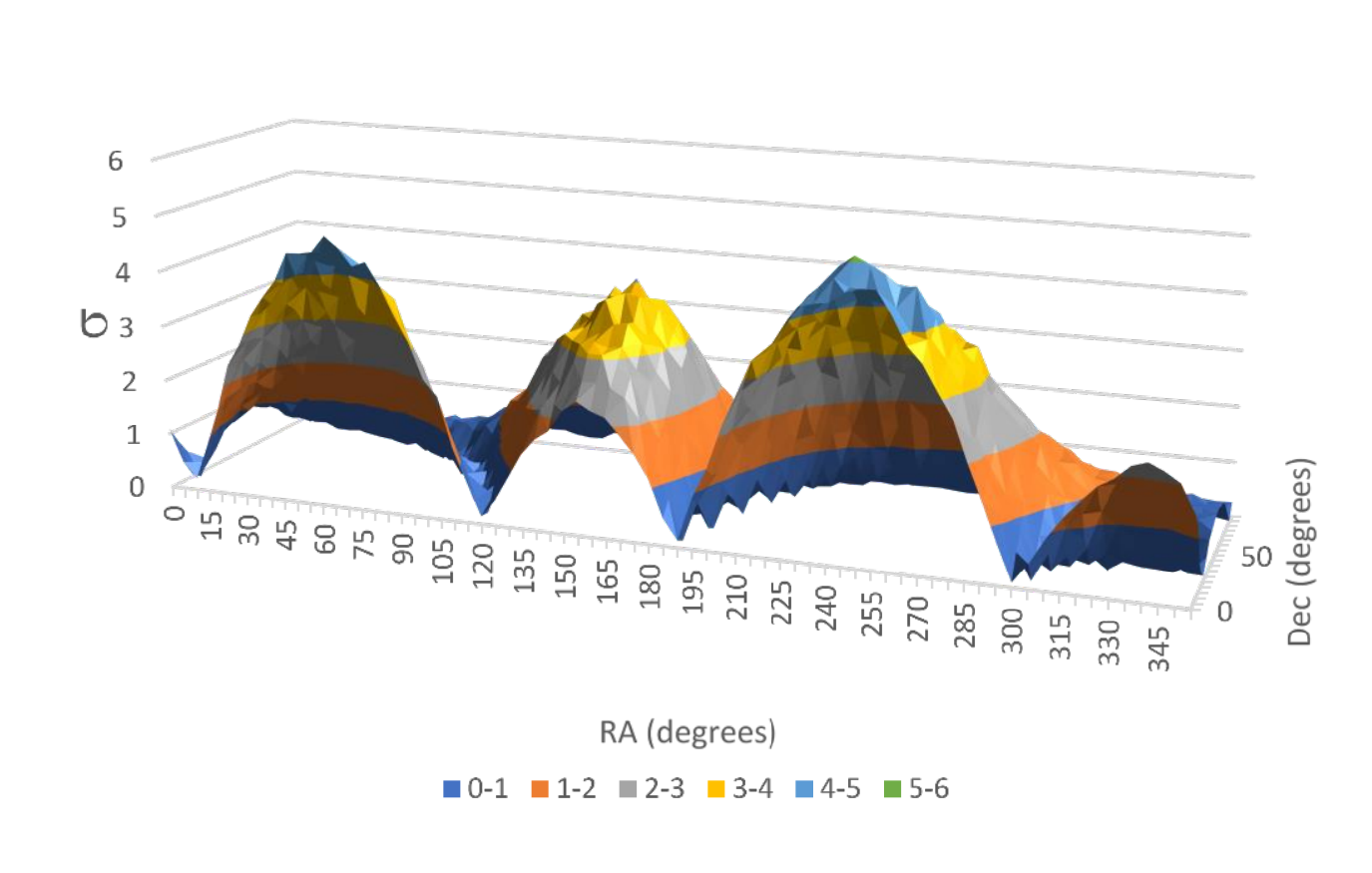}
\caption{The $\sigma$ of the likelihood of a octopole axis in different $(\alpha,\delta)$ combinations.}
\label{axis_quadrupole_z_015}
\end{figure}

Tables~\ref{dis_120_210} through~\ref{dis_30_300} show certain evidence that the asymmetry increases with the redshift. When fitting the quadrupole using just galaxies with $(z>0.15)$, the most likely axes are at $(\alpha=239^o,\delta=7^o)$ and $(\alpha=155^o,\delta=34^o)$ roughly at the same location as when using the entire dataset, but the $\sigma$ of the most likely axes is much higher with 5.94 and 8.16, respectively. Figure~\ref{axis_quadrupole_z_015} shows the $\sigma$ in all possible $(\alpha,\delta)$ combinations.

\subsection{Photometric asymmetry}
\label{photometric_differences}

The asymmetry between the number of clockwise and counterclockwise galaxies peaks in the RA range of $(120^o,210^o)$, which agrees with the photometric differences between clockwise and counterclockwise galaxies, that also peaks at around that part of the sky as reported in detail in \citep{shamir2017colour,shamir2017photometric,shamir2017large}. % The asymmetry is not significant in the RA ranges of $(210^o,300^o)$ and $(30^o,120^o)$, as also shown in \citep{shamir2017colour,shamir2017photometric,shamir2017large} for the photometric asymmetry. 
Some evidence for asymmetry can be identified in the RA range of $( > 300^o < 30^o)$. However, that RA range does not have a high population of galaxies in the dataset, making it difficult to profile a clear pattern.

Differences between the magnitude of clockwise and counterclockwise galaxies can also lead to differences in the number of detected clockwise and counterclockwise galaxies. Brighter galaxies are expected to be easier to detect and to identify its morphology, whether the morphology is analyzed manually or automatically. If a certain type of galaxies is brighter than the other, more galaxies of that type can be identified, leading to a higher galaxy count of that type. Therefore, if face-on clockwise galaxies at a certain part of the sky have, on average, a different magnitude than a face-on counterclockwise galaxy, that can lead to a difference in the number of clockwise and counterclockwise galaxies that are detected. On the other hand, galaxies that are closer to Earth tend to have lower apparent magnitude (brighter) than galaxies located deeper in the universe. Therefore, asymmetry between the number of clockwise and counterclockwise galaxies in a certain redshift range can lead to differences in the mean magnitude of these two sets of galaxies. It is therefore important to analyze the relationship between the differences in galaxy population and differences in magnitude.

Table~\ref{r_magnitude_120_210} shows the mean exponential r magnitude of the clockwise and counterclockwise galaxies in different redshift ranges in the RA range of  $(120^o,210^o)$. Flag magnitude values such as ``-999'' that are common in SDSS were ignored from the analysis, as they do not represent actual measured photometry. As the table shows, while clockwise galaxies tend to be brighter on average, there are no statistically significant differences between the exponential r magnitude in the clockwise galaxies and the exponential r magnitude in the counterclockwise galaxies. 

\begin{table}
{
%\footnotesize
\scriptsize
\begin{tabular}{|l|c|c|c|}
\hline
z   &   cw &  ccw & P (t-test) \\      
\hline
0-0.05     &  16.434$\pm$0.027 & 16.391$\pm$0.027	 & 0.26  \\
0.05-0.1 &	16.923$\pm$0.01  &  16.924$\pm$0.01 &  0.96 \\
0.1-0.15 &	 17.374$\pm$0.008 &  17.39$\pm$0.008 &  0.165 \\
0.15-0.2 &	17.684$\pm$0.014  & 17.693$\pm$0.012  & 	0.63  \\
0.2 - & 18.838$\pm$0.016  & 18.866$\pm$0.015  & 0.24 \\
% Total & 17.317$\pm$0.009 &   17.356$\pm$0.009 &  0.0027 \\
\hline
\end{tabular}
\caption{The mean exponential r magnitude of the clockwise and counterclockwise galaxies in different redshift ranges in the RA range of $(120^o,210^o)$.}
\label{r_magnitude_120_210}
}
\end{table}

\begin{figure*}[h]
\includegraphics[scale=0.8]{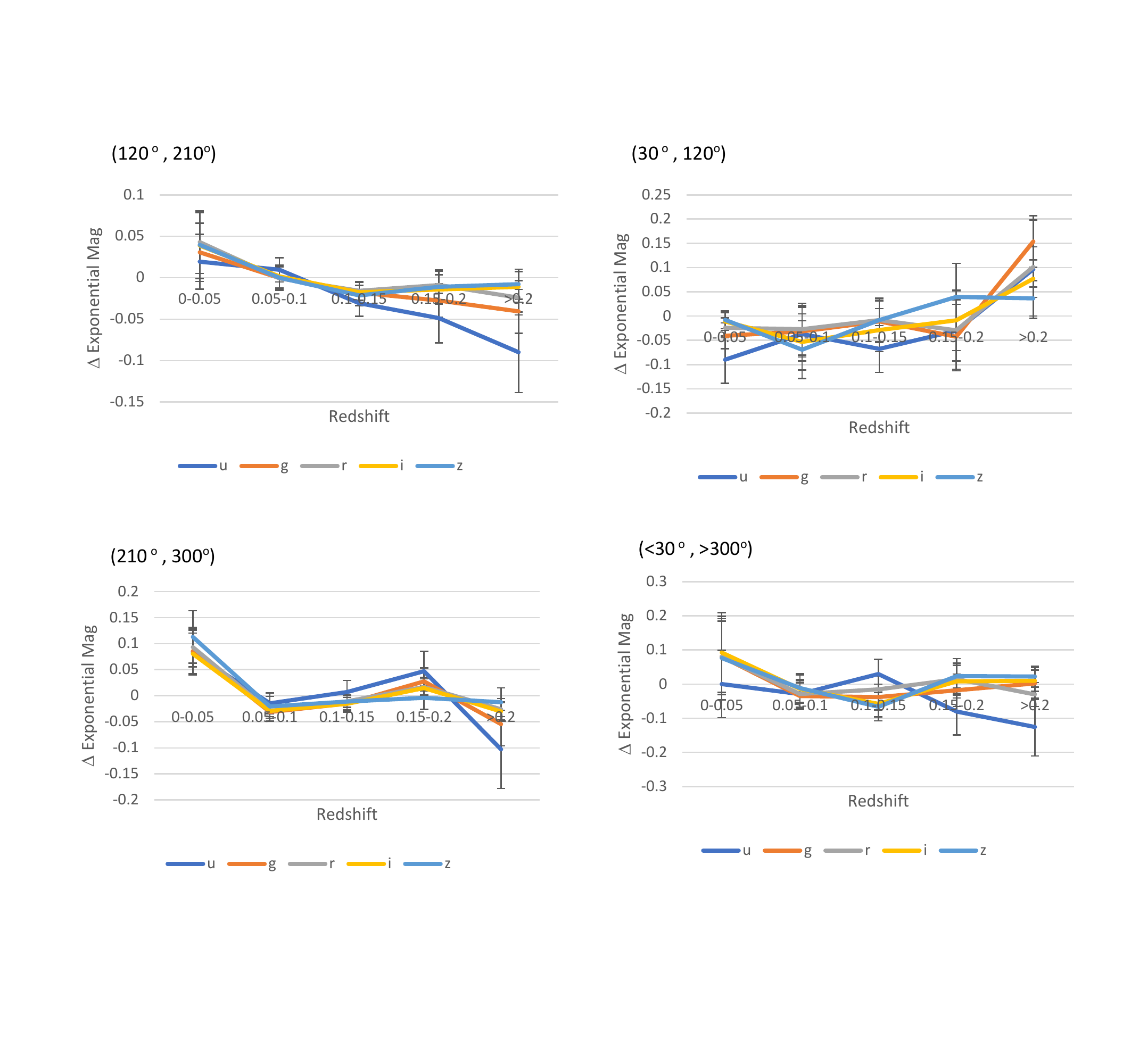}
\caption{The difference in magnitude (cw-ccw) in different redshift ranges in the four different RA ranges.}
\label{magnitude_all}
\end{figure*}

Table~\ref{magnitude_120_210} shows the differences between the magnitude of clockwise and counterclockwise galaxies in different bands and different redshift ranges. The values show an increase in the difference as the redshift gets higher, but even without correction for multiple tests none of these differences is statistically significant. Figure~\ref{magnitude_all} shows the difference in mean magnitude (cw-ccw) in different redshift ranges in the four RA ranges. None of the tests shows statistical significance. The only exception is the redshift range 0-0.05 in the RA range of (210$^o$,300$^o$), where the g and z band show statistical significance of $\sim$0.046 and $\sim$0.024, respectively. When correcting for the P values for the number of different redshift ranges, the q value of the z band is $\sim$0.12. 

\begin{table*}
{
%\footnotesize
\scriptsize
\begin{tabular}{|l|c|c|c|c|c|}
\hline
band & 0-0.05	& 0.05-0.1	& 0.1-0.15	& 0.15-0.2	& $>$0.2 \\
\hline
u	& 0.0193$\pm$0.033	& 0.0094$\pm$0.014	& -0.0308$\pm$0.016  & -0.04866$\pm$0.030 & -0.0899$\pm$0.049 \\
g	& 0.031$\pm$0.035	& -0.0004$\pm$0.013	& -0.018$\pm$0.012	& -0.0276$\pm$0.020	& -0.0406$\pm$0.027 \\
r	& 0.0429$\pm$0.038	& -0.0007$\pm$0.014	& -0.0158$\pm$0.011	& -0.0089$\pm$0.019 & -0.028$\pm$0.021\\
i	& 0.0388$\pm$0.040	& 0.0015$\pm$0.014	& -0.0176$\pm$0.012  & -0.0141$\pm$0.018	& -0.01102$\pm$0.019 \\
z	& 0.0395$\pm$0.041	& 0.00001$\pm$0.015  &	-0.0213$\pm$0.012 & -0.0109$\pm$0.019 & -0.0079$\pm$0.018 \\
\hline
\end{tabular}
\caption{The differences between the mean exponential magnitude of the clockwise and counterclockwise galaxies (cw-ccw) in different redshift ranges in the RA range of $(120^o,210^o)$.}
\label{magnitude_120_210}
}
\end{table*}

\subsection{Absolute magnitude}
\label{absolute_magnitude}

% SELECT top 10 objid,ra,dec,modelmag_r,modelmag_r-5*log10(4.28E+08*z) as abs_mag_r

As shown in Section~\ref{photometric_differences}, no significant differences in the photometry of clockwise and counterclockwise galaxies were identified. However, when comparing the apparent magnitude of two sets of galaxies such that the galaxies in one set are brighter than the galaxies in the other set, more galaxies from the brighter class might pass a certain threshold that makes them identifiable. For instance, let L and R be two identical galaxies, with the exception that L is a clockwise galaxy and R is a counterclockwise galaxy. Let galaxy L be slightly brighter than galaxy R, and galaxy L is just slightly above the threshold that allows identification of the galaxy morphology. Since L is just above the threshold that allows identification it will be identified and consequently included in the dataset and affect the results. R, however, will be slightly dimmer, and its morphology might not be identified and therefore it might be excluded from the dataset. That can lead to an increased population of L galaxies among the dimmer galaxies in the dataset. The higher population of dimmer galaxies can shift the mean magnitude of these galaxies.

If clockwise galaxies are completely symmetric to counterclockwise galaxies, the population and magnitude of all clockwise and counterclockwise galaxies should be identical within statistical error. However, as discussed earlier in this section, as well as in previous work \citep{shamir2013color,hoehn2014characteristics,shamir2016asymmetry,shamir2017colour,shamir2017photometric,shamir2017large}, differences between the number and magnitude of clockwise and counterclockwise galaxies have been observed, providing evidence of a certain asymmetry driven by the different spin patterns of galaxies.

Table~\ref{absolute_magnitude_120_210} shows the difference in the absolute magnitude of clockwise and counterclockwise galaxies in the RA range of $(120^o,210^o)$. As the table shows, the differences in the r, i, and z band are statistically significant, showing that counterclockwise galaxies in the dataset in that part of the sky are brighter than clockwise galaxies in the same sky region. Table~\ref{absolute_magnitude_30_300} shows that in the corresponding RA range in the opposite hemisphere $(<30^o, >300^o)$ the asymmetry in magnitude is inverse, but the photometric differences in that RA range are not statistically significant. It should be noted that the range $(<30^o, >300^o)$ contains far less galaxies in the dataset, and the low number increases the standard error and reduces the statistical power.

Tables~\ref{absolute_magnitude_30_120} and~\ref{absolute_magnitude_210_300} show the absolute magnitude differences in the different bands in the RA ranges of $(30^o,120^o)$ and $(210^o,300^o)$, respectively. In the RA range $(30^o,120^o)$ the clockwise galaxies are brighter than the counterclockwise galaxies, while in the RA range $(210^o,300^o)$ the counterclockwise galaxies are brighter. However, as Figure~\ref{distribution_ra} shows, the distribution of galaxies is not uniform in these RA ranges. The RA range of $(30^o,60^o)$ contains much more galaxies than $(90^o,120^o)$, which can explain a shift of the photometric asymmetry towards the asymmetry observed in $(<30^o, >300^o)$. In the RA range $(210^o,300^o)$ the number of galaxies is much higher in $(240^o,270^o)$, which could shift the asymmetry towards the hemisphere where the absolute magnitude of counterclockwise galaxies is brighter.

\begin{table}
{
%\footnotesize
\scriptsize
\begin{tabular}{|l|c|c|c|}
\hline
band   &   cw &  ccw & P (t-test) \\  
\hline
u  &  -18.627$\pm$0.010 & -18.637$\pm$0.010	 & 0.493  \\
g &	-19.929$\pm$0.009  &  -19.948$\pm$0.009 &  0.16 \\
r &	-20.687$\pm$0.01 &  -20.727$\pm$0.01 &  0.004\\
i &	-21.058$\pm$0.011  &-21.105$\pm$0.01  & 0.002  \\
z & -21.265$\pm$0.011  & -21.314$\pm$0.011  & 0.001 \\
\hline
\end{tabular}
\caption{The mean absolute exponential magnitude of the clockwise and counterclockwise galaxies in different redshift ranges in the RA range of $(120^o,210^o)$.}
\label{absolute_magnitude_120_210}
}
\end{table}

\begin{table}
{
%\footnotesize
\scriptsize
\begin{tabular}{|l|c|c|c|}
\hline
band   &   cw &  ccw & P (t-test) \\  
\hline
u  &  -18.527$\pm$0.023 & -18.494$\pm$0.023	 & 0.289  \\
g &	-19.791$\pm$0.019  &  -19.794$\pm$0.019 &  0.919 \\
r &	-20.603$\pm$0.021 &  -20.619$\pm$0.021 &  0.578\\
i &	-21.003$\pm$0.022  &-21.030$\pm$0.022  & 0.376  \\
z & -21.226$\pm$0.023  & -21.255$\pm$0.024  & 0.376 \\
\hline
\end{tabular}
\caption{The mean absolute exponential magnitude of the clockwise and counterclockwise galaxies in different redshift ranges in the RA range of $(<30^o, >300^o)$.}
\label{absolute_magnitude_30_300}
}
\end{table}

\begin{table}
{
%\footnotesize
\scriptsize
\begin{tabular}{|l|c|c|c|}
\hline
band   &   cw &  ccw & P (t-test) \\  
\hline
u  &  -18.463$\pm$0.03 & -18.372$\pm$0.028	 & 0.026  \\
g &	-19.718$\pm$0.024  &  -19.646$\pm$0.025 &  0.038 \\
r &	-20.554$\pm$0.027 &  -20.458$\pm$0.028 &  0.012 \\
i &	-21.967$\pm$0.028  &-21.855$\pm$0.029  & 0.006  \\
z & -21.205$\pm$0.029  & -21.091$\pm$0.030  & 0.007  \\
\hline
\end{tabular}
\caption{The mean absolute exponential magnitude of the clockwise and counterclockwise galaxies in different redshift ranges in the RA range of $(30^o,120^o)$.}
\label{absolute_magnitude_30_120}
}
\end{table}

\begin{table}
{
%\footnotesize
\scriptsize
\begin{tabular}{|l|c|c|c|}
\hline
band   &   cw &  ccw & P (t-test) \\  
\hline
u  &  -18.606$\pm$0.013 & -18.615$\pm$0.013	 & 0.604  \\
g &	-19.960$\pm$0.011  &  -19.964$\pm$0.011 &  0.797 \\
r &	-20.716$\pm$0.012 &  -20.728$\pm$0.012 &  0.475 \\
i &	-21.096$\pm$0.013  &-21.105$\pm$0.013  & 0.624  \\
z & -21.300$\pm$0.014  & -21.318$\pm$0.013  & 0.346 \\
\hline
\end{tabular}
\caption{The mean absolute exponential magnitude of the clockwise and counterclockwise galaxies in different redshift ranges in the RA range of $(210^o,300^o)$.}
\label{absolute_magnitude_210_300}
}
\end{table}

\begin{figure*}[h]
\includegraphics[scale=0.8]{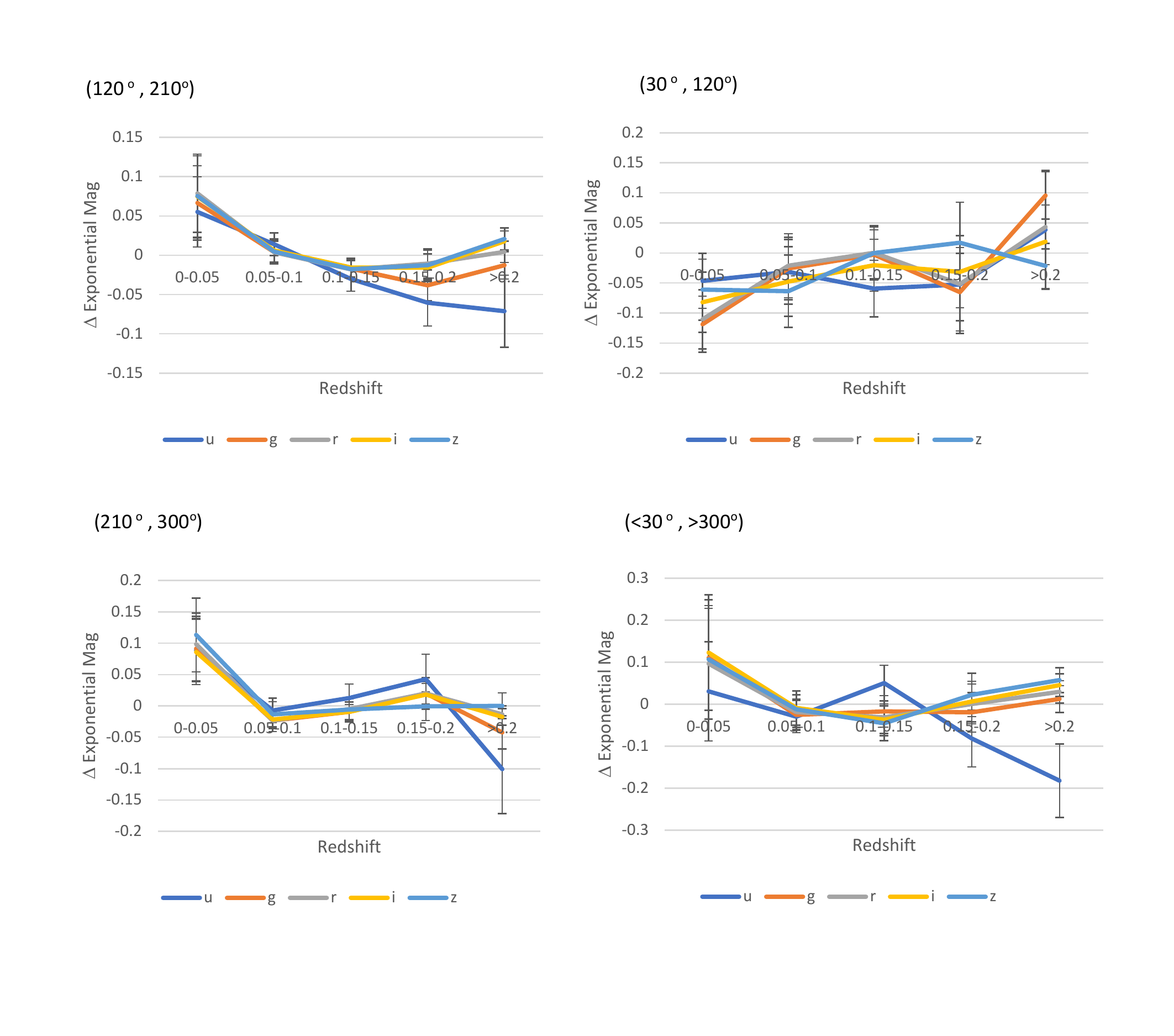}
\caption{The difference in absolute exponential magnitude (cw-ccw) in different redshift ranges in the four different RA ranges.}
\label{absolute_magnitude_all}
\end{figure*}

Figure~\ref{absolute_magnitude_all} shows the difference in mean absolute exponential magnitude (cw-ccw) in different redshift ranges in the four RA ranges. Like with the apparent magnitude, separating to redshift ranges does not reveal statistically significant differences between the magnitude of clockwise and counterclockwise galaxies.

% Table~\ref{limiting_magnitude_g_120_210} shows the differences in absolute magnitude (cw-ccw) in the g band. 

% \begin{table}
% {
% %\footnotesize
% \scriptsize
% \begin{tabular}{|l|c|c|c|}
% \hline
% g   &   cw &  ccw & P (t-test) \\  
% \hline
% $<$16.5  &  -19.974$\pm$0.028 & -20.035$\pm$0.025	 & 0.104  \\
% 16.5-17.5 &	-19.999$\pm$0.015  &  -20.026$\pm$0.014 &  0.18 \\
% 17.5-18.5 &	-20.012$\pm$0.009 &  -20.031$\pm$0.009 &  0.1355 \\
% 18.5-19.5 &	-20.031$\pm$0.008  &-20.047$\pm$0.008  & 0.1573  \\
% \hline
% \end{tabular}
% \caption{The mean absolute exponential magnitude of the clockwise and counterclockwise galaxies in different apparent magnitude ranges in the RA range of $(210^o,300^o)$.}
% \label{limiting_magnitude_g_120_210}
% }
% \end{table}

% \begin{figure*}[h]
% \includegraphics[scale=0.8]{limiting_magnitude.pdf}
% \caption{The difference in absolute exponential magnitude (cw-ccw) with different apparent magnitude ranges in the four different RA ranges.}
% \label{limiting_magnitude}
% \end{figure*}

\subsection{Comparison with normalized redshift}
\label{normalize}

As discussed in Section~\ref{photometric_differences}, the differences in the number of clockwise and counterclockwise galaxies in different redshift ranges can be due to differences in the magnitude of clockwise and counterclockwise galaxies. For instance, if counterclockwise galaxies are brighter than clockwise galaxies at a certain part of the sky, it is expected that more counterclockwise galaxies will be detected, and therefore that difference in magnitude can exhibit itself in the form of an increased population of clockwise galaxies. On the other hand, if the ratio between clockwise and counterclockwise changes with the redshift, that would exhibit itself in different mean magnitudes.

To test the link between redshift, magnitude and spin direction asymmetry, a set of galaxies in the RA range of $(120^o,180^o)$, where the asymmetry peaks, was selected. Each clockwise galaxy in the dataset was paired with a counterclockwise galaxy such that the redshift difference between the two galaxies was less than 0.001. That led to two equal-sized datasets such that one contained galaxies with clockwise spin patterns and the other contained galaxies with counterclockwise spin patterns. Due to the selection of the galaxies, the mean redshift of both datasets was almost identical, as well as the redshift distribution. Tables~\ref{norm_abs_g} and~\ref{norm_abs_i} show the mean absolute exponential magnitude in the g and i bands, respectively. The tables show that when the redshift distribution is similar in both sets of galaxies, the absolute magnitude is not statistically significant.

\begin{table*}
{
%\footnotesize
\scriptsize
\begin{tabular}{|l|c|c|c|c|c|c|c|}
\hline
Redshift   &   mean g cw  &  mean g ccw & mean z cw  & mean z ccw & count cw & count ccw & P (t-test) \\  
\hline
0-0.1    &  -19.477$\pm$0.018 & -19.505$\pm$0.018 &	0.060$\pm$0.0003 &  0.060$\pm$0.0003    & 5807 & 5807 & 0.27 \\
0.1-0.2 &	-20.546$\pm$0.01  &  -20.532$\pm$0.01     & 0.135$\pm$0.0004    & 0.135$\pm$0.0004 &3724  & 3724 & 0.32 \\
$>$0.2 &	-20.177$\pm$0.020 &  -20.196$\pm$0.020   &  0.333$\pm$0.002 &  0.333$\pm$0.002     & 1521 & 1521 & 0.5 \\
Total   &	-19.927$\pm$0.012  & -19.939$\pm$0.012   &   0.123$\pm$0.0009 &  0.123$\pm$0.0009 & 11052 & 11052 &  0.48 \\
\hline
\end{tabular}
\caption{The mean absolute g magnitude of the clockwise and counterclockwise galaxies in different redshift ranges. The galaxies are in the RA range of $(120^o,180^o)$ and paired such that each clockwise galaxy in the dataset is matched by a counterclockwise galaxy with a very close redshift ($\Delta<0.001$).}
\label{norm_abs_g}
}
\end{table*}

\begin{table*}
{
%\footnotesize
\scriptsize
\begin{tabular}{|l|c|c|c|c|c|c|c|}
\hline
Redshift   &   mean g cw &  mean g ccw & mean z cw  & mean z ccw & count cw & count ccw & P (t-test) \\  
\hline
0-0.1    &  -20.356$\pm$0.02 & -20.388$\pm$0.02 &	0.060$\pm$0.0003 &  0.060$\pm$0.0003    & 5807 & 5807 & 0.26 \\
0.1-0.2 &	-21.680$\pm$0.01  &  -21.673$\pm$0.01     & 0.135$\pm$0.0004    & 0.135$\pm$0.0004 &3724  & 3724 & 0.62 \\
$>$0.2 &	-22.218$\pm$0.013 &  -22.240$\pm$0.013   &  0.333$\pm$0.002 &  0.333$\pm$0.002     & 1521 & 1521 & 0.23 \\
Total   &	-21.059$\pm$0.014  & -21.076$\pm$0.014   &   0.123$\pm$0.0009 &  0.123$\pm$0.0009 & 11052 & 11052 & 0.39 \\
\hline
\end{tabular}
\caption{The mean absolute i magnitude of the clockwise and counterclockwise galaxies in different redshift ranges. The galaxies are in the RA range of $(120^o,180^o)$ and paired such that each clockwise galaxy in the dataset is matched by a counterclockwise galaxy with a very close redshift ($\Delta<0.001$).}
\label{norm_abs_i}
}
\end{table*}

Tables~\ref{norm_g} and~\ref{norm_i} show the differences in the apparent magnitude in the g and i bands when the redshift distribution in the two datasets is normalized. Both tables show no statistical significant differences between the magnitude.

\begin{table*}
{
%\footnotesize
\scriptsize
\begin{tabular}{|l|c|c|c|c|c|c|c|}
\hline
Redshift   &   g cw &  g ccw & z cw  & z ccw & count cw & count ccw & P (t-test) \\  
\hline
0-0.1    &  17.337$\pm$0.014 & 17.310$\pm$0.014 &	0.060$\pm$0.0003 &  0.060$\pm$0.0003    & 5807 & 5807 & 0.17\\
0.1-0.2 &	18.226$\pm$0.01  &  18.240$\pm$0.01     & 0.135$\pm$0.0004    & 0.135$\pm$0.0004 &3724  & 3724 & 0.32 \\
$>$0.2 &	20.580$\pm$0.026 &  20.566$\pm$0.026   &  0.333$\pm$0.002 &  0.333$\pm$0.002     & 1521 & 1521 & 0.7 \\
Total   &	17.978$\pm$0.012  & 17.967$\pm$0.012   &   0.123$\pm$0.0008 &  0.123$\pm$0.0008 & 11052 & 11052 & 0.76 \\
\hline
\end{tabular}
\caption{The mean exponential apparent g magnitude of the clockwise and counterclockwise galaxies in different redshift ranges. The galaxies are in the RA range of $(120^o,180^o)$ and paired such that each clockwise galaxy in the dataset is matched by a counterclockwise galaxy with a very close redshift ($\Delta<0.001$).}
\label{norm_g}
}
\end{table*}

\begin{table*}
{
%\footnotesize
\scriptsize
\begin{tabular}{|l|c|c|c|c|c|c|c|}
\hline
Redshift   &   g cw &  g ccw & z cw  & z ccw & count cw & count ccw & P (t-test) \\  
\hline
0-0.1    &  16.458$\pm$0.015 & 16.428$\pm$0.015 &	0.060$\pm$0.0003 &  0.060$\pm$0.0003       & 5807 & 5807 & 0.16 \\
0.1-0.2 &	17.091$\pm$0.009  &  17.099$\pm$0.009     & 0.135$\pm$0.0004    & 0.135$\pm$0.0004 & 3724 & 3724 & 0.53 \\
$>$0.2 &	18.488$\pm$0.017 &  18.465$\pm$0.017   &  0.333$\pm$0.002 &  0.333$\pm$0.002        & 1521 & 1521 & 0.38 \\
Total   &	16.951$\pm$0.011  & 16.935$\pm$0.011   &   0.123$\pm$0.0009 &  0.123$\pm$0.0009    & 11052 & 11052 & 0.3 \\
\hline
\end{tabular}
\caption{The mean exponential i magnitude of the clockwise and counterclockwise galaxies in different redshift ranges. The galaxies are in the RA range of $(120^o,180^o)$ and paired such that each clockwise galaxy in the dataset is matched by a counterclockwise galaxy with a very close redshift ($\Delta<0.001$).}
\label{norm_i}
}
\end{table*}

While the absolute magnitude shows statistically significant differences between clockwise and counterclockwise galaxies in the initial dataset of galaxies, the absolute magnitude does not show any statistically significant differences when the redshift distribution is normalized. That might indicate that the observed differences in magnitude are the result of the redshift differences between clockwise and counterclockwise galaxies.

Tables~\ref{norm_absolute_u} through ~\ref{norm_absolute_i} show a similar analysis, but instead of normalizing the distribution of the two datasets by the redshift, the two datasets are normalized by the absolute magnitude. That is, the difference between the mean redshift of clockwise and counterclockwise galaxies is measured such that  the distribution of the absolute magnitude in the dataset of clockwise galaxies is similar to the distribution of the absolute magnitude in the dataset of counterclockwise galaxies. The tables show that when the distribution of the absolute magnitude in both datasets is similar, the difference between the mean redshift in both datasets is statistically significant. 

A similar experiment was done with the apparent magnitude. Tables~\ref{norm_apparent_g} through~\ref{norm_apparent_z} show the same analysis when the distribution of the apparent magnitude of the clockwise galaxies is similar to the distribution of the apparent magnitude of the counterclockwise galaxies. Unlike with the absolute magnitude, when the distribution of the apparent magnitude is similar in both datasets the statistical significant is not as clear, especially in the g band. However, it can be noticed that when the number of galaxies is higher, the statistical becomes stronger.

\begin{table*}
{
%\footnotesize
\scriptsize
\begin{tabular}{|l|c|c|c|c|c|c|c|}
\hline
u & z cw &  z ccw &  u cw & u ccw & cw count & ccw count & P (t-test)  \\
\hline
-20 - -21 & 0.1770$\pm$0.0029 & 0.1855$\pm$0.0032 & -20.299$\pm$0.009 & -20.298$\pm$0.009 & 677 & 677 & 0.05 \\ 
-20- -19 & 0.1244$\pm$0.0011 & 0.1293$\pm$0.0012 & -19.427$\pm$0.005 & -19.427$\pm$0.005 & 3459 & 3459 & 0.003 \\ 
-19 - -18 &  0.1024$\pm$0.0011 & 0.1089$\pm$0.0012 & -18.543$\pm$0.005 & -18.543$\pm$0.005 & 3834 & 3834 & $<$0.0001\\
-18 - -17 & 0.0879$\pm$0.0020 & 0.0954$\pm$0.0022 & -17.601$\pm$0.007 & -17.601$\pm$0.007 & 1645 & 1645  & 0.012 \\
Total & 0.1131$\pm$0.0007 & 0.1194$\pm$0.0008 & -18.823$\pm$0.008 & -18.823$\pm$0.008 & 9615 & 9615 & $<$0.0001\\
\hline
\end{tabular}
\caption{The mean redshift when the two galaxy datasets have similar distribution of the absolute u magnitude. The galaxies are in the RA range of $(120^o,180^o)$ and paired such that each clockwise galaxy in the dataset is matched by a counterclockwise galaxy with similar magnitude ($\Delta<0.01$).}
\label{norm_absolute_u}
}
\end{table*}

\begin{table*}
{
%\footnotesize
\scriptsize
\begin{tabular}{|l|c|c|c|c|c|c|c|}
\hline
g & z cw &  z ccw &  g cw & g ccw & cw count & ccw count & P (t-test)  \\
\hline
% -22- -21 & 0.1471$\pm$0.002 & 0.1475$\pm$0.002 & -21.259$\pm$0.006 & -21.259$\pm$0.006 & 1085 & 1085 & 0.880  \\

-22- -21 & 0.1463$\pm$0.0017 & 0.1456$\pm$0.0017 & -21.255$\pm$0.006 & -21.259$\pm$0.006 & 1085 & 1085 & 0.77  \\
-21- -20 & 0.1365$\pm$0.001 & 0.1450$\pm$0.001 & -20.457$\pm$0.004 & -20.456$\pm$0.004 & 5305 & 5305 & $<$0.0001  \\
-20- -19 & 0.1268$\pm$0.002 & 0.1350$\pm$0.002 & -19.615$\pm$0.005 & -19.616$\pm$0.005 & 3405 & 3405 & 0.005  \\
-19- -18 & 0.0621$\pm$0.003 & 0.0664$\pm$0.003 & -18.606$\pm$0.010 & -18.606$\pm$0.010 & 800 & 800 & 0.250  \\
Total & 0.1252$\pm$0.001 & 0.1323$\pm$0.001 & -19.985$\pm$0.010 & -19.985$\pm$0.010 & 10595 & 10595 & $<$0.0001  \\
\hline
\end{tabular}
\caption{The mean redshift when the two galaxy datasets have similar distribution of the absolute g magnitude. The galaxies are in the RA range of $(120^o,180^o)$ and paired such that each clockwise galaxy in the dataset is matched by a counterclockwise galaxy with similar magnitude ($\Delta<0.01$).}
\label{norm_absolute_g}
}
\end{table*}

\begin{table*}
{
%\footnotesize
\scriptsize
\begin{tabular}{|l|c|c|c|c|c|c|c|}
\hline
i & z cw &  z ccw &   i cw &  i ccw &  cw count & ccw count & P (t-test)  \\
\hline
-23- -22 & 0.2352$\pm$0.0024 & 0.2446$\pm$0.0024 & -22.3155$\pm$0.005 & -22.3164$\pm$0.005 & 2442 & 2442 & 0.005  \\
-22- -21 & 0.1197$\pm$0.0010 & 0.1229$\pm$0.0011 & -21.5189$\pm$0.004 & -21.5196$\pm$0.004 & 4646 & 4646 & 0.026  \\
-21- -20 & 0.0721$\pm$0.0007 & 0.0728$\pm$0.0007 & -20.58847$\pm$0.006 & -20.5876$\pm$0.006 & 2435 & 2435 & 0.51  \\
Total & 0.1260$\pm$0.0009 & 0.1293$\pm$0.0010 & -21.1489$\pm$0.012 & -21.1491$\pm$0.012 & 9523 & 9523 & 0.013  \\
\hline
\end{tabular}
\caption{The mean redshift when the galaxy dataset is normalized by the absolute i magnitude. The galaxies are in the RA range of $(120^o,180^o)$ and paired such that each clockwise galaxy in the dataset is matched by a counterclockwise galaxy with similar magnitude ($\Delta<0.01$).}
\label{norm_absolute_i}
}
\end{table*}

\begin{table*}
{
%\footnotesize
\scriptsize
\begin{tabular}{|l|c|c|c|c|c|c|c|}
\hline
u & z cw &  z ccw &  u cw &  u ccw &  cw count & ccw count & P (t-test)  \\
\hline
17-18 & 0.0522$\pm$0.0012 & 0.0558$\pm$0.0014 & 17.614$\pm$0.008 & 17.614$\pm$0.008 & 1049 & 1049 & 0.05 \\
18-19 & 0.0806$\pm$0.0009 & 0.0810$\pm$0.0008 & 18.578$\pm$0.005 & 18.578$\pm$0.005 & 3118 & 3118 & 0.73 \\
19-20 & 0.1153$\pm$0.0009 & 0.1181$\pm$0.0009 & 19.441$\pm$0.005 & 19.441$\pm$0.005 & 3811 & 3811 & 0.027 \\
20-21 & 0.1527$\pm$0.0021 & 0.1563$\pm$0.0021 & 20.391$\pm$0.008 & 20.391$\pm$0.008 & 1240 & 1240 & 0.225 \\
\hline
\end{tabular}
\caption{The mean redshift when the two galaxy datasets have similar distribution of the apparent u magnitude. The galaxies are in the RA range of $(120^o,180^o)$ and paired such that each clockwise galaxy in the dataset is matched by a counterclockwise galaxy with similar magnitude ($\Delta<0.01$).}
\label{norm_apparent_u}
}
\end{table*}

\begin{table*}
{
%\footnotesize
\scriptsize
\begin{tabular}{|l|c|c|c|c|c|c|c|}
\hline
g & z cw &  z ccw &  g cw &  g ccw &  cw count & ccw count & P (t-test)  \\
\hline
16-17 & 0.0544$\pm$0.001 & 0.0547$\pm$0.001 & 16.588$\pm$0.007 & 16.589$\pm$0.007 & 1470 & 1470 & 0.71  \\
17-18 & 0.0826$\pm$0.0005 & 0.0843$\pm$0.0006 & 17.568$\pm$0.005 & 17.568$\pm$0.005 & 3854 & 3854  & 0.016\\
18-19 & 0.1247$\pm$0.0007 & 0.1256$\pm$0.0007 & 18.367$\pm$0.004 & 18.368$\pm$0.004 & 3461 & 3461 & 0.36 \\
19-20 & 0.2067$\pm$0.0031 & 0.2027$\pm$0.0033 & 19.421$\pm$0.016 & 19.421$\pm$0.015 & 383 & 383  & 0.36 \\
% Total & 0.1242$\pm$0.001 & 0.1255$\pm$0.001 & 18.094$\pm$0.013 & 18.095$\pm$0.013 & 9271 & 9271 & 0.31  \\
\hline
\end{tabular}
\caption{The mean redshift when the two galaxy datasets have similar distribution of the apparent g magnitude. The galaxies are in the RA range of $(120^o,180^o)$ and paired such that each clockwise galaxy in the dataset is matched by a counterclockwise galaxy with similar magnitude ($\Delta<0.01$).}
\label{norm_apparent_g}
}
\end{table*}

\begin{table*}
{
%\footnotesize
\scriptsize
\begin{tabular}{|l|c|c|c|c|c|c|c|}
\hline
r & z cw &  z ccw &  r cw &  r ccw &  cw count & ccw count & P (t-test)  \\
\hline
15-16 & 0.0471$\pm$0.0007 & 0.0485$\pm$0.0008 & 15.630$\pm$0.009 & 15.630$\pm$0.009 & 849 & 849 & 0.18 \\
16-17 & 0.0744$\pm$0.0006 & 0.0754$\pm$0.0006 & 16.585$\pm$0.005 & 16.585$\pm$0.005 & 2768 & 2768 & 0.09 \\
17-18 & 0.1117$\pm$0.0006 & 0.1133$\pm$0.0006 & 17.485$\pm$0.004 & 17.485$\pm$0.004 & 5333 & 5333 & 0.0265 \\
18-19 & 0.2354$\pm$0.0033 & 0.2371$\pm$0.0032 & 18.465$\pm$0.013 & 18.465$\pm$0.013 & 633 & 633 &  0.71 \\
\hline
\end{tabular}
\caption{The mean redshift when the two galaxy datasets have similar distribution of the apparent r magnitude. The galaxies are in the RA range of $(120^o,180^o)$ and paired such that each clockwise galaxy in the dataset is matched by a counterclockwise galaxy with similar magnitude ($\Delta<0.01$).}
\label{norm_apparent_r}
}
\end{table*}

\begin{table*}
{
%\footnotesize
\scriptsize
\begin{tabular}{|l|c|c|c|c|c|c|c|}
\hline
i & z cw &  z ccw &  i cw &  i ccw &  cw count & ccw count & P (t-test)  \\
\hline
16-17 & 0.0880$\pm$0.0006 & 0.0899$\pm$0.0006 & 16.575$\pm$0.004 & 16.575$\pm$0.004 & 3921 & 3921 & 0.025  \\
17-18 & 0.1241$\pm$0.0009 & 0.1267$\pm$0.0009 & 17.323$\pm$0.003 & 17.325$\pm$0.003 & 3927 & 3927 & 0.05  \\
18-19 & 0.3307$\pm$0.0027 & 0.3282$\pm$0.0027 & 18.560$\pm$0.009 & 18.560$\pm$0.009 & 936 & 936 & 0.51  \\
19-20 & 0.3701$\pm$0.0063 & 0.3853$\pm$0.0059 & 19.276$\pm$0.013 & 19.276$\pm$0.013 & 346 & 346 & 0.09  \\
% Total & 0.1252$\pm$0.0009 & 0.1269$\pm$0.0009 & 16.967$\pm$0.010 & 16.967$\pm$0.010 & 9130 & 9130 & 0.18 \\
\hline
\end{tabular}
\caption{The mean redshift when the two galaxy datasets have similar distribution of the apparent i magnitude. The galaxies are in the RA range of $(120^o,180^o)$ and paired such that each clockwise galaxy in the dataset is matched by a counterclockwise galaxy with similar magnitude ($\Delta<0.01$).}
\label{norm_apparent_i}
}
\end{table*}

\begin{table*}
{
%\footnotesize
\scriptsize
\begin{tabular}{|l|c|c|c|c|c|c|c|}
\hline
z & redshift cw &  redshift ccw &  z cw &  z ccw &  cw count & ccw count & P (t-test)  \\
\hline
15 - 16  & 0.0621$\pm$0.0006 & 0.0629$\pm$0.0006 & 15.596$\pm$0.007 & 15.596$\pm$0.007 & 1689 & 1689 & 0.34 \\
16-17 & 0.0987$\pm$0.0006 & 0.1003$\pm$0.0006 & 16.571$\pm$0.004 & 16.572$\pm$0.004 & 4734 & 4734 & 0.05 \\
17 - 18 & 0.1381$\pm$0.0012 & 0.1414$\pm$0.0012 & 17.285$\pm$0.004 & 17.285$\pm$0.004 & 2786 & 2786 & 0.05 \\
18-19 & 0.3138$\pm$0.0032 & 0.3208$\pm$0.0028 & 18.373$\pm$0.009 & 18.373$\pm$0.009 & 658 & 658 & 0.09 \\ 
\hline
\end{tabular}
\caption{The mean redshift when the two galaxy datasets have similar distribution of the apparent z magnitude. The galaxies are in the RA range of $(120^o,180^o)$ and paired such that each clockwise galaxy in the dataset is matched by a counterclockwise galaxy with similar magnitude ($\Delta<0.01$).}
\label{norm_apparent_z}
}
\end{table*}

\section{Conclusion}
\label{conclusion}

Previous studies showed evidence of differences between galaxies with clockwise and counterclockwise spin patterns \citep{shamir2013color,hoehn2014characteristics,shamir2016asymmetry,shamir2017colour,shamir2017photometric,shamir2017large}. While some previous attempts were limited by the bias of the human eye \citep{land2008galaxy}, model-driven automatic morphological analysis of galaxies was able to produce much larger datasets that are not biased by the human perception, showing strong evidence that the distribution of clockwise and counterclockwise galaxies as seen from Earth is not uniform in all parts of the sky, and correspond to the direction of observation \citep{shamir2012handedness}.

Here a dataset of $\sim6.4\cdot10^4$ galaxies with spectra is used to show an asymmetry in the distribution of clockwise and counterclockwise galaxies as observed by SDSS. The entire process of data analysis is machine-based, and therefore no human bias can affect the results. Moreover, not only that the data analysis process is computer-based, none of the stages of data analytics involves machine learning. Therefore, no human bias or other effects can lead to bias in the training set, as no training set is used in any stages of the process. The process of data analysis is therefore deterministic, does not rely on any human input, and cannot be biased due to imbalanced or biased training data.

The analysis of the data shows that the number of clockwise galaxies is different from the number counterclockwise galaxies observed from Earth by SDSS, and that difference changes based on the direction of observation. The difference also changes with the redshift, showing that at least in one part of the sky the number of counterclockwise galaxies increases compared to clockwise galaxies as the redshift gets higher. The spin pattern of a spiral galaxy is also an indication of its spin direction \citep{iye2019spin}, and therefore the results show that in the redshift range that can be observed by SDSS (with identifiable galaxy morphology) the earlier universe is more homogeneous in terms of the distribution of galaxy spin directions.

Fitting the distribution of the spin directions reveals multipole alignment, with best fit to quadrupole alignment. The large-scale quadrupole alignment has been observed by CMB data \citep{cline2003does,gordon2004low,zhe2015quadrupole}, and led to cosmological theories that shift from the standard cosmological models \citep{feng2003double,piao2004suppressing,rodrigues2008anisotropic,piao2005possible,jimenez2007cosmology,bohmer2008cmb}. It also led to the ellipsoidal universe model \citep{campanelli2006ellipsoidal,campanelli2007cosmic,gruppuso2007complete}, and new geometrical models of the universe that can explain the quadrupole alignment \citep{weeks2004well,efstathiou2003low}. As the spin pattern of a spiral galaxy is also an indication of its spin direction \citep{iye2019spin}, the distribution of the spin directions in such a large scale can be an indication of a rotating universe \citep{godel1949example,ozsvath1962finite,ozsvath2001approaches}.

Naturally, redshift and photometry are highly correlated, and the ability to identify the morphology of a galaxy depends on its brightness. Therefore, asymmetry between the photometry of clockwise and counterclockwise galaxies can lead to a different number of detected galaxies in different redshift ranges. Also, asymmetry in the distribution of clockwise and counterclockwise galaxies in different redhsift ranges can lead to differences in the photometry as measured from Earth, as a higher population of a certain type of galaxies in the higher redshift ranges will lead to a higher (dimmer) apparent magnitude of that type. To profile that, I used one dataset in which clockwise and counterclockwise galaxies have very similar distribution of the redshift, and another dataset in which clockwise and counterclockwise galaxies have very similar distribution of the absolute magnitude. When the distribution of the redshift is similar for both types of galaxies, no statistically significant differences can be identified between the absolute magnitude of clockwise galaxies and the absolute magnitude of counterclockwise galaxies. However, when the distribution of the absolute magnitude is similar in both datasets, the differences in the redshift are still detected and are still statistically significant. That provides evidence that the difference in the population of galaxies is not driven by magnitude differences that allow for better detection of one type of galaxy over another.

The spin pattern of a galaxy is a crude measurement, and there is no known atmospheric or other effect that can make a clockwise galaxy seem counterclockwise or vice versa. The classification of the galaxies was done in a fully automatic manner, and no human bias can have any impact on the results. The Ganalyzer algorithm used for the classification is a model-driven algorithm that uses clear rules for its classification. It is not driven by a machine learning process, in which the training data can add bias to the result, and the complex nature of the rules that determine the output makes it difficult to assess the way the classification is made. The asymmetry is identified in different parts of the sky that are in opposite hemispheres. Previous work showed with very strong statistical significance that the asymmetry between clockwise and counterclockwise galaxies changes with the direction of observation \citep{shamir2017colour,shamir2017photometric,shamir2017large}. These results had very good agreement across two different sky surveys: SDSS and PanSTARRS \citep{shamir2017large}. That observation further eliminates the possibility of a software error, as such error should have exhibited itself in the form of consistent asymmetry in all parts of the sky, regardless of the direction of observation. As Table~\ref{directions} shows, the direction of asymmetry is different in different parts of the sky, such that some parts show a higher number of clockwise galaxies, and other parts of the sky show a higher number of counterclockwise galaxies. A computer error would have expected to exhibit itself in the form of a consistent bias, and would not invert in different parts of the sky, that are also in different hemispheres. Tables~\ref{dis_120_210} through~\ref{dis_30_300} show that the asymmetry increases with the redshift, and the asymmetry grows in opposite directions in different parts of the sky. That also indicates that the asymmetry is not the result of a software error, as a software error would be consistent in all parts of the sky, and would lead to a decreasing asymmetry with the redshift due to the increasing difficulty in classifying high redshift galaxies that tend to be smaller and dimmer than galaxies with low redshift.

As Figure~\ref{distribution} shows, the span of the galaxies used in this study is far larger than any known astrophysical structure, and might therefore provide evidence for violation of the isotropy assumption of the universe as observed from Earth. While the isotropy and homogeneity assumptions are pivotal to standard cosmological models, these assumptions have not been proven, and spatial homogeneity cannot be verified directly \citep{ellis1979homogeneity}. Some evidence of isotropy violation at the cosmological scale have been observed through other messengers such as gamma ray bursts \citep{meszaros2019oppositeness} and cosmic microwave background \citep{aghanim2014planck,hu1997cmb,cooray2003cosmic,ben2012parity,eriksen2004asymmetries}. The distribution of spin directions shown here provides evidence for the violation of the cosmological isotropy and homogeneity assumptions, and a spatial structure at a cosmological scale. Future and more powerful sky surveys such as LSST, or combination of optical surveys such as the DECam Legacy Survey \citep{blum2016decam} with powerful spectroscopy surveys such as DESI \citep{font2014desi} can provide a higher resolution profiling of the asymmetry.

% http://www.astro.ucla.edu/~wright/CMB-dipole-history.html
% https://www.nature.com/articles/222971a0

\section*{Acknowledgments}
% This study was supported in part by NSF grants AST-1903823 and IIS-1546079.

Funding for the Sloan Digital Sky Survey IV has been provided by the Alfred P. Sloan Foundation, the U.S. Department of Energy Office of Science, and the Participating Institutions. SDSS-IV acknowledges support and resources from the Center for High-Performance Computing at the University of Utah. The SDSS web site is www.sdss.org.

SDSS-IV is managed by the Astrophysical Research Consortium for the Participating Institutions of the SDSS Collaboration including the Brazilian Participation Group, the Carnegie Institution for Science, Carnegie Mellon University, the Chilean Participation Group, the French Participation Group, Harvard-Smithsonian Center for Astrophysics, Instituto de Astrof\'isica de Canarias, The Johns Hopkins University, Kavli Institute for the Physics and Mathematics of the Universe (IPMU) / 
University of Tokyo, the Korean Participation Group, Lawrence Berkeley National Laboratory, Leibniz Institut f\"ur Astrophysik Potsdam (AIP), Max-Planck-Institut f\"ur Astronomie (MPIA Heidelberg), Max-Planck-Institut f\"ur Astrophysik (MPA Garching), Max-Planck-Institut f\"ur Extraterrestrische Physik (MPE), National Astronomical Observatories of China, New Mexico State University, New York University, University of Notre Dame, Observat\'ario Nacional / MCTI, The Ohio State University, Pennsylvania State University, Shanghai Astronomical Observatory, United Kingdom Participation Group, Universidad Nacional Aut\'onoma de M\'exico, University of Arizona, University of Colorado Boulder, University of Oxford, University of Portsmouth, University of Utah, University of Virginia, University of Washington, University of Wisconsin, Vanderbilt University, and Yale University.

\bibliographystyle{apalike}

%\bibliography{myref}

\bibliography{Liorshamir_ms}

\begin{thebibliography}{}

\bibitem[Aghanim et~al., 2014]{aghanim2014planck}
Aghanim, N., Armitage-Caplan, C., Arnaud, M., Ashdown, M., Atrio-Barandela, F.,
  Aumont, J., Baccigalupi, C., Banday, A., Barreiro, R., Bartlett, J., et~al.
  (2014).
\newblock Planck 2013 results. xxvii. doppler boosting of the cmb: Eppur si
  muove.
\newblock {\em Astronomy \& Astrophysics}, 571:A27.

\bibitem[Ben-David et~al., 2012]{ben2012parity}
Ben-David, A., Kovetz, E.~D., and Itzhaki, N. (2012).
\newblock Parity in the cosmic microwave background: Space oddity.
\newblock {\em The Astrophysical Journal}, 748(1):39.

\bibitem[Blum et~al., 2016]{blum2016decam}
Blum, R.~D., Burleigh, K., Dey, A., Schlegel, D.~J., Meisner, A.~M., Levi, M.,
  Myers, A.~D., Lang, D., Moustakas, J., Patej, A., et~al. (2016).
\newblock The decam legacy survey.
\newblock In {\em American Astronomical Society Meeting Abstracts\# 228},
  volume 228.

\bibitem[Bohmer and Mota, 2008]{bohmer2008cmb}
Bohmer, C.~G. and Mota, D.~F. (2008).
\newblock Cmb anisotropies and inflation from non-standard spinors.
\newblock {\em Physics Letters B}, 663(3):168--171.

\bibitem[Campanelli et~al., 2006]{campanelli2006ellipsoidal}
Campanelli, L., Cea, P., and Tedesco, L. (2006).
\newblock Ellipsoidal universe can solve the cosmic microwave background
  quadrupole problem.
\newblock {\em Physical review letters}, 97(13):131302.

\bibitem[Campanelli et~al., 2007]{campanelli2007cosmic}
Campanelli, L., Cea, P., and Tedesco, L. (2007).
\newblock Cosmic microwave background quadrupole and ellipsoidal universe.
\newblock {\em Physical Review D}, 76(6):063007.

\bibitem[Cline et~al., 2003]{cline2003does}
Cline, J.~M., Crotty, P., and Lesgourgues, J. (2003).
\newblock Does the small cmb quadrupole moment suggest new physics?
\newblock {\em Journal of Cosmology and Astroparticle Physics}, 2003(09):010.

\bibitem[Cooray et~al., 2003]{cooray2003cosmic}
Cooray, A., Melchiorri, A., and Silk, J. (2003).
\newblock Is the cosmic microwave background circularly polarized?
\newblock {\em Physics Letters B}, 554(1-2):1--6.

\bibitem[Dojcsak and Shamir, 2014]{dojcsak2014quantitative}
Dojcsak, L. and Shamir, L. (2014).
\newblock Quantitative analysis of spirality in elliptical galaxies.
\newblock {\em New Astronomy}, 28:1--8.

\bibitem[Efstathiou, 2003]{efstathiou2003low}
Efstathiou, G. (2003).
\newblock Is the low cosmic microwave background quadrupole a signature of
  spatial curvature?
\newblock {\em Monthly Notices of the Royal Astronomical Society},
  343(4):L95--L98.

\bibitem[Ellis, 1979]{ellis1979homogeneity}
Ellis, G. (1979).
\newblock The homogeneity of the universe.
\newblock {\em General Relativity and Gravitation}, 11(4):281--289.

\bibitem[Eriksen et~al., 2004]{eriksen2004asymmetries}
Eriksen, H.~K., Hansen, F.~K., Banday, A.~J., G{\'o}rski, K.~M., and Lilje,
  P.~B. (2004).
\newblock Asymmetries in the cosmic microwave background anisotropy field.
\newblock {\em The Astrophysical Journal}, 605(1):14.

\bibitem[Feng and Zhang, 2003]{feng2003double}
Feng, B. and Zhang, X. (2003).
\newblock Double inflation and the low cmb quadrupole.
\newblock {\em Physics Letters B}, 570(3-4):145--150.

\bibitem[Font-Ribera et~al., 2014]{font2014desi}
Font-Ribera, A., McDonald, P., Mostek, N., Reid, B.~A., Seo, H.-J., and Slosar,
  A. (2014).
\newblock Desi and other dark energy experiments in the era of neutrino mass
  measurements.
\newblock {\em Journal of Cosmology and Astroparticle Physics}, 2014(05):023.

\bibitem[G{\"o}del, 1949]{godel1949example}
G{\"o}del, K. (1949).
\newblock An example of a new type of cosmological solutions of einstein's
  field equations of gravitation.
\newblock {\em Reviews of modern physics}, 21(3):447.

\bibitem[Gordon and Hu, 2004]{gordon2004low}
Gordon, C. and Hu, W. (2004).
\newblock Low cmb quadrupole from dark energy isocurvature perturbations.
\newblock {\em Physical Review D}, 70(8):083003.

\bibitem[Gruppuso, 2007]{gruppuso2007complete}
Gruppuso, A. (2007).
\newblock Complete statistical analysis for the quadrupole amplitude in an
  ellipsoidal universe.
\newblock {\em Physical Review D}, 76(8):083010.

\bibitem[Hoehn and Shamir, 2014]{hoehn2014characteristics}
Hoehn, C. and Shamir, L. (2014).
\newblock Characteristics of clockwise and counterclockwise spiral galaxies.
\newblock {\em AN}, 335(2):189--192.

\bibitem[Hu and White, 1997]{hu1997cmb}
Hu, W. and White, M. (1997).
\newblock A cmb polarization primer.
\newblock {\em arXiv preprint astro-ph/9706147}.

\bibitem[Iye and Sugai, 1991]{iye1991catalog}
Iye, M. and Sugai, H. (1991).
\newblock A catalog of spin orientation of southern galaxies.
\newblock {\em The Astrophysical Journal}, 374:112--116.

\bibitem[Iye et~al., 2019]{iye2019spin}
Iye, M., Tadaki, K., and Fukumoto, H. (2019).
\newblock Spin parity of spiral galaxies i--corroborative evidence for trailing
  spirals.
\newblock {\em arXiv preprint arXiv:1910.10926}.

\bibitem[Jim{\'e}nez and Maroto, 2007]{jimenez2007cosmology}
Jim{\'e}nez, J.~B. and Maroto, A.~L. (2007).
\newblock Cosmology with moving dark energy and the cmb quadrupole.
\newblock {\em Physical Review D}, 76(2):023003.

\bibitem[Kuminski and Shamir, 2016]{kuminski2016computer}
Kuminski, E. and Shamir, L. (2016).
\newblock A computer-generated visual morphology catalog of ~3,000,000 sdss
  galaxies.
\newblock {\em The Astrophysical Journal Supplement Series}, 223(2):20.

\bibitem[Land et~al., 2008]{land2008galaxy}
Land, K., Slosar, A., Lintott, C., Andreescu, D., Bamford, S., Murray, P.,
  Nichol, R., Raddick, M.~J., Schawinski, K., Szalay, A., et~al. (2008).
\newblock Galaxy zoo: the large-scale spin statistics of spiral galaxies in the
  sloan digital sky survey.
\newblock {\em Monthly Notices of the Royal Astronomical Society},
  388(4):1686--1692.

\bibitem[Lee et~al., 2019]{lee2019mysterious}
Lee, J.~H., Pak, M., Song, H., Lee, H.-R., Kim, S., and Jeong, H. (2019).
\newblock Mysterious coherence in several-megaparsec scales between galaxy
  rotation and neighbor motion.
\newblock {\em The Astrophysical Journal}, 884(2):104.

\bibitem[Longo, 2011]{longo2011detection}
Longo, M.~J. (2011).
\newblock Detection of a dipole in the handedness of spiral galaxies with
  redshifts z~ 0.04.
\newblock {\em Physics Letters B}, 699(4):224--229.

\bibitem[M{\'e}sz{\'a}ros, 2019]{meszaros2019oppositeness}
M{\'e}sz{\'a}ros, A. (2019).
\newblock An oppositeness in the cosmology: Distribution of the gamma ray
  bursts and the cosmological principle.
\newblock {\em AN}, 340(7):564--569.

\bibitem[Ozsv{\'a}th and Sch{\"u}cking, 1962]{ozsvath1962finite}
Ozsv{\'a}th, I. and Sch{\"u}cking, E. (1962).
\newblock Finite rotating universe.
\newblock {\em Nature}, 193(4821):1168--1169.

\bibitem[Ozsvath and Sch{\"u}cking, 2001]{ozsvath2001approaches}
Ozsvath, I. and Sch{\"u}cking, E. (2001).
\newblock Approaches to g{\"o}del's rotating universe.
\newblock {\em Classical and Quantum Gravity}, 18(12):2243.

\bibitem[Piao, 2005]{piao2005possible}
Piao, Y.-S. (2005).
\newblock Possible explanation to a low cmb quadrupole.
\newblock {\em Physical Review D}, 71(8):087301.

\bibitem[Piao et~al., 2004]{piao2004suppressing}
Piao, Y.-S., Feng, B., and Zhang, X. (2004).
\newblock Suppressing the cmb quadrupole with a bounce from the contracting
  phase to inflation.
\newblock {\em Physical Review D}, 69(10):103520.

\bibitem[Puerari et~al., 1997]{puerari1997relative}
Puerari, I., Garcia-Gomez, C., and Garijo, A. (1997).
\newblock On the relative orientation of binary galaxies.
\newblock {\em arXiv preprint astro-ph/9709236}.

\bibitem[Rodrigues, 2008]{rodrigues2008anisotropic}
Rodrigues, D.~C. (2008).
\newblock Anisotropic cosmological constant and the cmb quadrupole anomaly.
\newblock {\em Physical Review D}, 77(2):023534.

\bibitem[Shamir, 2011a]{shamir2011ganalyzer}
Shamir, L. (2011a).
\newblock Ganalyzer: A tool for automatic galaxy image analysis.
\newblock {\em ApJ}, 736(2):141.

\bibitem[Shamir, 2011b]{ganalyzer_ascl}
Shamir, L. (2011b).
\newblock Ganalyzer: A tool for automatic galaxy image analysis.
\newblock {\em The Astrophysics Source Code Library}, page ascl:1105.011.

\bibitem[Shamir, 2012]{shamir2012handedness}
Shamir, L. (2012).
\newblock Handedness asymmetry of spiral galaxies with z< 0.3 shows cosmic
  parity violation and a dipole axis.
\newblock {\em Physics Letters B}, 715(1-3):25--29.

\bibitem[Shamir, 2013]{shamir2013color}
Shamir, L. (2013).
\newblock Color differences between clockwise and counterclockwise spiral
  galaxies.
\newblock {\em Galaxies}, 1(3):210--215.

\bibitem[Shamir, 2016]{shamir2016asymmetry}
Shamir, L. (2016).
\newblock Asymmetry between galaxies with clockwise handedness and
  counterclockwise handedness.
\newblock {\em ApJ}, 823(1):32.

\bibitem[Shamir, 2017a]{shamir2017colour}
Shamir, L. (2017a).
\newblock Colour asymmetry between galaxies with clockwise and counterclockwise
  handedness.
\newblock {\em ApSS}, 362(2):33.

\bibitem[Shamir, 2017b]{shamir2017large}
Shamir, L. (2017b).
\newblock Large-scale photometric asymmetry in galaxy spin patterns.
\newblock {\em PASA}, 34:e44.

\bibitem[Shamir, 2017c]{shamir2017photometric}
Shamir, L. (2017c).
\newblock Photometric asymmetry between clockwise and counterclockwise spiral
  galaxies in sdss.
\newblock {\em PASA}, 34:e011.

\bibitem[Sofue, 1992]{sofue1992spins}
Sofue, Y. (1992).
\newblock Spins of interacting galaxies and a'tri-axial'angular momentum
  hypothesis.
\newblock {\em Publications of the Astronomical Society of Japan}, 44:L1--L5.

\bibitem[Weeks et~al., 2004]{weeks2004well}
Weeks, J., Luminet, J.-P., Riazuelo, A., and Lehoucq, R. (2004).
\newblock Well-proportioned universes suppress the cosmic microwave background
  quadrupole.
\newblock {\em Monthly Notices of the Royal Astronomical Society},
  352(1):258--262.

\bibitem[Zhe et~al., 2015]{zhe2015quadrupole}
Zhe, C., Xin, L., and Sai, W. (2015).
\newblock Quadrupole-octopole alignment of cmb related to the primordial power
  spectrum with dipolar modulation in anisotropic spacetime.
\newblock {\em Chinese Physics C}, 39(5):055101.

\end{thebibliography}

\end{document}